\pdfoutput=1
\documentclass[a4paper,fleqn,usenatbib]{mnras}

\usepackage{newtxtext,newtxmath}

\usepackage[T1]{fontenc}
\usepackage{ae,aecompl}

\usepackage{graphicx}	
\usepackage{amsmath}	
\usepackage{amssymb}	

\usepackage{verbatim}

\usepackage{etoolbox}
\makeatletter
\patchcmd\@combinedblfloats{\box\@outputbox}{\unvbox\@outputbox}{}{%
   \errmessage{\noexpand\@combinedblfloats could not be patched}%
}%
 \makeatother

\title[The life cycle of circumnuclear gas discs]{The life cycle of starbursting circumnuclear gas discs}

\author[M.~Schartmann et al.]
  {M.~Schartmann,$^{1,2,3}$\thanks{E-mail: mschartmann@swin.edu.au}
   J.~Mould,$^{1}$
   K.~Wada,$^{4}$
   A.~Burkert,$^{2,3}$
   M.~Durr\'{e},$^{1}$
   M.~Behrendt,$^{2,3}$
   \newauthor
   R.~I.~Davies,$^{3}$
   L.~Burtscher$^{3,5}$
\\
$^{1}$Centre for Astrophysics and Supercomputing, Swinburne University
of Technology, P.O. Box 218, Hawthorn, Victoria 3122, Australia\\
$^{2}$Universit\"ats-Sternwarte M\"unchen, Scheinerstra\ss e 1, D-81679 M\"unchen, Germany\\
$^{3}$Max-Planck-Institut f\"ur extraterrestrische Physik, Postfach 1312, Giessenbachstr., D-85741 Garching, Germany\\ 
$^{4}$Graduate School of Science and Engineering, Kagoshima University, Kagoshima 890-0065, Japan\\
$^{5}$Sterrewacht Leiden, Leiden University, Niels-Bohr-Weg 2, 2300 CA Leiden, Netherlands\\
}

\date{Accepted XXX. Received YYY; in original form ZZZ}

\pubyear{2017}

\volume{{\rm in press}}

\begin{document}
\label{firstpage}
\pagerange{\pageref{firstpage}--\pageref{lastpage}}
\maketitle

\begin{abstract}
High-resolution observations from the sub-mm to the optical wavelength regime
resolve the central 
few 100~pc region of nearby galaxies in great detail. They reveal a large diversity of features: 
thick gas and stellar discs, nuclear 
starbursts, in- and outflows, central activity, jet interaction, etc. 
Concentrating on the role circumnuclear discs play in the life cycles of galactic nuclei,
we employ 3D adaptive mesh refinement hydrodynamical simulations with the
{\sc Ramses} code to self-consistently trace the evolution from a quasi-stable gas disc,
undergoing gravitational (Toomre) instability, the formation of clumps and stars and
the disc's subsequent, partial dispersal via stellar feedback.
Our approach builds upon the observational finding that many nearby Seyfert galaxies have undergone  
intense nuclear starbursts in their recent past and in many nearby sources star formation is concentrated in
a handful of clumps on a few 100\,pc distant from the galactic centre. 
We show that such observations can be understood as the result of 
gravitational instabilities in dense circumnuclear discs.
By comparing these simulations to available integral field unit observations of a sample
of nearby galactic nuclei, we find consistent gas and stellar masses, kinematics, star 
formation and outflow properties. 
Important ingredients in the simulations are the self-consistent treatment of star formation
and the dynamical evolution of the stellar distribution as well as the modelling of a 
delay time distribution for the supernova feedback.
The knowledge of the resulting simulated density structure 
and kinematics on pc scale is vital for understanding inflow and feedback processes towards galactic scales. 
\end{abstract}

\begin{keywords}
hydrodynamics -- galaxies: evolution -- galaxies: ISM -- galaxies: kinematics and dynamics -- galaxies: nuclei -- galaxies: starburst
\end{keywords}





\section{Introduction}
\label{sec:introduction}

Circumnuclear gas discs are ubiquitously observed in the central regions of many 
classes of galaxies. Interacting systems and   
Ultra-luminous Infrared Galaxies \citep[ULIRGS,][]{Downes_98,Medling_14} have 
especially attracted the interest of many researchers. The detected discs in these
systems typically 
have dimensions of several hundred parsecs, gas masses of the order of 
$10^8$ to $10^{10}\,$M$_\odot$ and ratios of rotational velocity to velocity dispersion
of $v/\sigma$ between 1 and 5 \citep{Medling_14}. It is found 
that a large fraction of the coexisting stellar discs are consistent 
with being formed recently (<30\,Myr) within the gaseous discs.
We will especially concentrate on the case of active galaxies, where we are interested in 
investigating the interplay between nuclear disc evolution and nuclear activity, as well
as their mutual relation. 
According to the so-called {\it Unified Scheme of Active Galactic Nuclei (AGN)} 
\citep{Miller_83,Antonucci_93,Urry_95} these objects are thought to be powered by 
accretion onto their central supermassive black hole. Angular momentum conservation of
the infalling gas leads to the formation of a viscously heated accretion disc.
The latter can easily outshine the stars of the whole galaxy and
illuminates a larger gas and dust reservoir on parsec scale, the
so-called {\it dusty, molecular torus}
\citep[e.~g.~][]{Krolik_88,Nenkova_02,Hoenig_06,Schartmann_08,Stalevski_12,Wada_12,Siebenmorgen_15,Wada_16}. 
Mass transfer onto and
through these structures is often provided from an adjacent circumnuclear disc
or mini-spiral \citep[e.~g.~][]{Prieto_05,Hicks_09,Davies_14,Durre_14} which
typically reaches out to several hundreds of parsecs in nearby 
galactic nuclei. These discs are found to be made up of a
multi-phase mixture of gas and dust at various temperatures and
various stages of ionisation arising from shocks, star formation (SF) and
the radiation from the active nucleus as well as stars. 
Additional frequently observed components are outflows, partly in the form
of collimated, highly powerful jets, but also in wide-angle, lower velocity,
but high mass loaded winds. The strengths of the various phenomena
differ from source to source. 
All of these components are at the limits of resolution, not just
of our largest telescopes and best instrumentation, but also of
hydrodynamical codes that deal with their time evolution.

Observationally, such systems have e.~g.~been investigated
by \citet{Davies_07}, \citet{Hicks_13} and \citet{Lin_16}, concentrating on a sample of nearby 
Seyfert galaxies. They find a much higher rate of circumnuclear discs in active
galaxies compared to their inactive sample. 
Such studies with integral field units at the largest available telescopes with resolutions of 
a few parsec find that gaseous and stellar structures 
are often cospatial and share similar kinematics, indicating that 
stars may have formed in-situ from the gas discs.
Morphologies range from smooth, star forming discs and mini-spirals \citep{Prieto_05} 
over star formation concentrated in clumps \citep{Durre_14} 
to very disturbed filamentary outflowing structures \citep{Durre_17}
and are readily visible in dust extinction maps as well \citep{Prieto_05,Prieto_14}.
Most of these sources do not show any signs of recent merging activity that could 
provide the necessary torques to transfer gas towards the central region. Hence the 
discs / mini-spirals are thought to be formed mostly by secular evolution processes
\citep{Orban_11,Maciejewski_04a,Maciejewski_04b},
driving gas into 
the nuclear region (typically up to several hundred parsecs), 
e.~g.~via bars \citep{Sakamoto_99,Sheth_05,Krumholz_15}.
After enough gas has been accumulated, the discs become gravitationally 
unstable \citep{Toomre_64,Behrendt_15}. This is the starting point of our simulations, 
which we approximate with idealised, marginally (Toomre) unstable discs.

Theoretically, the formation and evolution of circumnuclear gas discs has mainly been 
studied
within simulations of mergers of gas-rich galaxies (in isolation and within cosmological 
frameworks). 
Gravitational torques are able to remove angular momentum from shocked interstellar 
gas, leading to infall. 
Cosmological zoom simulations are nowadays able to follow these processes down to the 
formation of 
circumnuclear discs similar to the ones observed on several 100 parsecs scale 
\citep{Levine_08,Hopkins_10}. Due to their violent past, many of these discs are warped 
and can have complex kinematics, partly 
decoupled from the rest of the galaxy \citep{Barnes_02}. These discs typically grow 
inside-out
and disc formation can take place over a long period of time, due to infalling tidal tails.
\citet{Roskar_15} find that directly after the merger, a strong starburst event evacuates 
the central region surrounding the supermassive black hole (SMBH), but a circumnuclear disc of several 100\,pc size
reforms on a time scale of roughly 10\,Myr. Most of these studies concentrate on the effect of 
such a gas disc on the evolution and the in-spiral of a black hole binary, following the recent merging event
\citep[e.~g.~][]{Chapon_13}.

The dynamical state of (starbursting) circumnuclear discs in nearby active galaxies has been studied by
\citet{Fukuda_00}, \citet{Wada_02} and \citet{Wada_09}. They find that supernova (SN) feedback under starburst conditions 
turns a rotationally supported thin disc into a turbulent, clumpy, geometrically thick toroidal structure 
on tens of parsecs scale with significant contributions to the total obscuration 
properties. The generated turbulence efficiently drives gas towards the SMBH. 
For the case of an already activated galactic nucleus, a {\it radiation-pressure driven fountain} flow -- enabled by 
the direct radiation pressure from a central source -- 
can also lead to a geometrically thick distribution and drive a significant outflow along the symmetry axis \citep{Wada_12,Wada_15}
and is able to describe some observed properties of nearby Seyfert galaxies \citep{Schartmann_14}. The combination of SN and central
radiation feedback enabled a good description of the observable properties of the very well-studied Circinus galaxy \citep{Wada_16}.

Our approach is complementary to this set of simulations. In this first publication, we start by characterising  
self-consistent self-gravitating hydrodynamical simulations of  
Toomre unstable circumnuclear discs. Spanning a full starburst cycle, we study their 
gravitational instability, star forming properties and the driving of winds towards 
galactic scales.
Such simulations -- constrained by the mentioned observations -- will allow us to derive a consistent picture
of the mass budget of these systems concerning star formation and in- and outflows driven by the starburst. 
The tens of parsecs scale vicinity of SMBHs is not only important for fuelling the central putative molecular, 
dusty torus \citep[e.~g.~][]{Schartmann_08,Schartmann_14}, but can also provide a fraction of the 
necessary obscuration \citep{Hicks_09,Prieto_14,Wada_02,Wada_15}, which in turn is important for assessing the efficiency 
of SMBH feedback towards galactic scales. 

We present our physical model, parameters and assumptions in Sect.~\ref{sec:physical_model}, discuss the 
evolution of its hydrodynamical realisation and of a parameter study in Sect.~\ref{sec:results} and 
compare the results to observations (Sect.~\ref{sec:comp_obs}). 
We critically discuss our obtained results in Sect.~\ref{sec:discussion}, before
we conclude our analysis in Sect.~\ref{sec:conclusions}.

\section{The physical model and its numerical representation}
\label{sec:physical_model}

In this first simulation series, we include only a subset of physical
processes and a simple initial condition guided by observations.
This will form the basis of our investigation of the life cycles 
and dynamics of gas and stars in circumnuclear discs in the nuclei 
of nearby active galaxies. Subsequently adding 
more physical processes will give us a detailed understanding
of their internal workings.

\subsection{The initial gas disc setup and the background potential}
\label{sec:bg_pot_ic_disc}

\begin{figure*}
  \includegraphics[width=\textwidth]{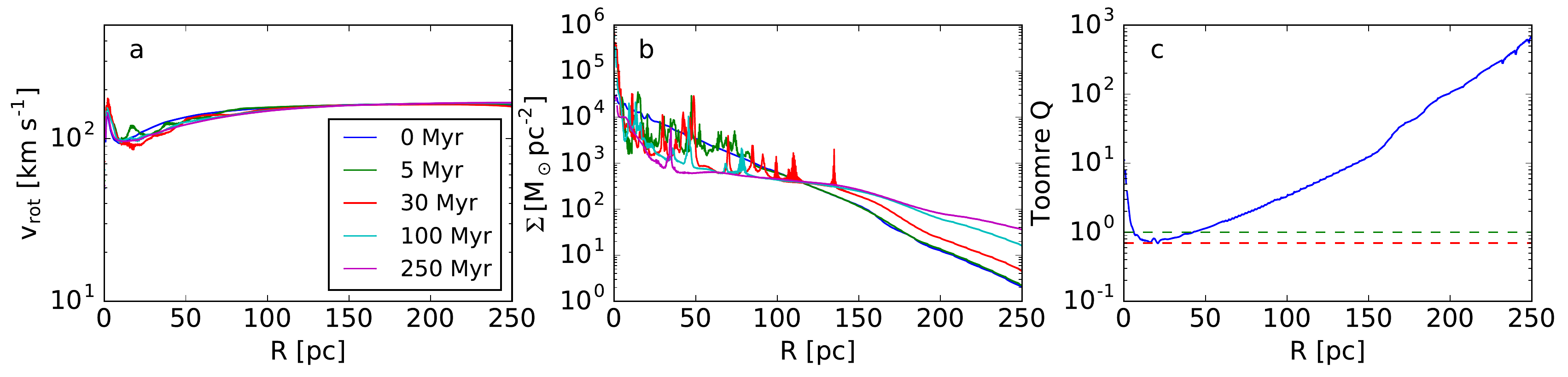}
  \caption{Azimuthally averaged rotation velocity (a), surface density (b) and Toomre Q parameter (c) of the disc at various evolutionary stages
   (see legend). In panel (c), only the initial condition is shown. The centre reaches high values of the Toomre Q parameter, 
   which is followed by a marginally unstable region (up to several ten parsecs distance from the centre) that
   forms stars and turns the initially smooth into a clumpy disc, whereas the outer region remains stable and smooth.}
  \label{fig:galnuc_radial_plots}
\end{figure*}

The most common feature of the IFU observations are nuclear, rotating disc
structures. Hence, in our basic model, we assume that there is a pre-existing gas
disc with a radially exponential surface density distribution with a scaling length
of 30\,pc, following the observations presented by \citet{Hicks_09}. It is rotationally supported in the radial direction and
in approximate vertical hydrostatic equilibrium with a background potential (BH and stellar bulge, see below) and the self-gravity
of the disc itself. 
The disc temperature is set to $T_{\mathrm{disc,ini}}=10^4$\,K. The latter is thought to 
replace an unresolved micro-turbulent pressure floor. 
Such a micro-turbulent pressure floor can be thought of as arising from the
transfer of gravitational potential energy from the accretion of gas towards the centre
\citep[][see discussion in Sect.~\ref{sec:lim_phys}]{Klessen_10}. 
In the limit of small disc masses around a point mass, this setup leads to a 
vertical Gaussian distribution of the gas density. In the self-gravity
limit when the gas disc dominates over the central point source,  
a $\sec^2$ density distribution results.
Our case is intermediate and partially dominated by the extended background potential 
and we numerically calculate the vertical structure of the disc.
It is surrounded by a hot atmosphere with $T_{\mathrm{atm,ini}}=10^6$\,K in approximate hydrostatic equilibrium 
with the BH, bulge and gas self-gravity potential.
Being interested in the local galaxy population, we set the central supermassive black hole mass 
to $M_{\mathrm{BH}}=10^7\,$M$_\odot$, which is implemented as a Gaussian 
background potential with a full width at half maximum (FWHM) of 2\,pc. 
The second component is the (old) stellar bulge that dominates the background potential. 
Its mass is set to $M_{\mathrm{bulge}}=8.5 \times 10^9\,$M$_\odot$, derived following the scaling relation between central 
supermassive black holes and their stellar bulge masses given by \citet{Haering_04}.
We model it with a spherical Hernquist \citep{Hernquist_90} potential with a half-mass radius of $r_\mathrm{h}^\mathrm{bulge}=820$\,pc. The latter 
has been approximated following \citet[][Fig.~3]{Berg_14}.
Together with the self-gravity of the gas, this background potential results in a flat rotation curve 
over most part of our computational domain (Fig.~\ref{fig:galnuc_radial_plots}a), as typically derived from observations of nearby 
AGN \citep[e.~g.~][]{Davies_07,Hicks_09}.
The size of the initial disc is chosen to
be a few times the scale length of the observed discs and similar to the 
field of view of the IFU and spectral observations
which will be used to compare our results to. 
The computational domain is a cube with a side length of 2048\,pc.
A summary of the model 
parameters is given in Table~\ref{tab:model_params}. 
\begin{table}
\caption[Model parameters of the simulations.]{Model
  parameters of the simulations.}
\label{tab:model_params}
\begin{tabular}{lrlr}
\hline
$M_{\mathrm{BH}}$ & $10^7\,$M$_{\odot}$ & $M_{\mathrm{gas,ini}}$ &  $10^8\,$M$_{\odot}$ \\
$M_{\mathrm{bulge}}$ & $8.5 \times 10^9\,\mathrm{M}_{\odot}$ & $r_\mathrm{h}^\mathrm{bulge}$ & 820\,pc \\
$T_{\mathrm{disc,ini}}$ &  $10^4$\,K & $T_{\mathrm{atm,ini}}$ &  $10^6$\,K  \\
$\Delta x_{\mathrm{box}}$ & 2048\,pc & $l_{\mathrm{disc}}$ & 30\,pc \\
$n_{\mathrm{SF}}$ & $2 \times 10^{6}\,\mathrm{cm}^{-3}$ & $\epsilon_\mathrm{SF}$ & 0.02\%\\
$\eta_\mathrm{SN}$ & 0.1 & $E_{\mathrm{SN}}$ & $10^{51}$\,erg \\
$m^*_{\mathrm{SN}}$ & 10\,M$_{\odot}$ & $m_*$ & 100\,M$_\odot$ \\
$m^{\mathrm{SN}}_{\mathrm{threshold}}$ & 1000\,M$_{\odot}$ & $R^{\mathrm{max}}_{\mathrm{bubble}}$ & 10\,pc\\
\hline
\end{tabular}
\newline
\medskip
$M_{\mathrm{BH}}$ is the central black hole mass, 
$M_{\mathrm{gas,ini}}$ is the total initial gas mass (disc plus atmosphere),
$M_{\mathrm{bulge}}$ is the mass of the (old) stellar bulge,
$r_\mathrm{h}^\mathrm{bulge}$ its half-mass radius,
$T_{\mathrm{disc,ini}}$ the initial temperature of the disc,
$T_{\mathrm{atm,ini}}$ the initial temperature of the atmosphere,
$\Delta x_{\mathrm{box}}$ the size of the computational domain,
$l_{\mathrm{disc}}$ the scale length of the exponential gas disc,
$n_{\mathrm{SF}}$  is the hydrogen number density threshold for star formation,
$\epsilon_\mathrm{SF}$ is the star formation efficiency per free-fall time,
$\eta_\mathrm{SN}$ is the fraction of stellar mass that goes into supernovae and
$E_{\mathrm{SN}}$ is the assumed energy injected per SN and
$m^*_{\mathrm{SN}}$ is the assumed mass of a SN progenitor star,
$m_*$ is the typical mass of a stellar population / star cluster,
$m^{\mathrm{SN}}_{\mathrm{threshold}}$ is the gas mass threshold to determine the 
SN bubble radius,
$R^{\mathrm{max}}_{\mathrm{bubble}}$ is the maximum SN bubble radius.\\
\end{table}
\subsection{Numerical hydrodynamics, self-gravity and adaptive mesh refinement}
\label{sec:numhydro}
We solve the system of hydrodynamical equations and the Poisson equation with the help of the 
{\sc Ramses} \citep{Teyssier_02} code, which uses a second-order Godunov-type hydrodynamical adaptive mesh refinement (AMR) scheme
with an octree-based data structure.
The solver proposed by \citet[][HLL Riemann solver]{Harten_83} is used to calculate the solution to the Riemann problem.
The complex physics of star formation (SF) and stellar feedback necessitate
a very high spatial resolution. 
We adopt a quasi-Lagrangian AMR strategy and refine a cell up to a maximum
refinement level whenever its mass (including the old stellar bulge and the central SMBH) 
exceeds a given threshold mass for the respective 
level. We furthermore require that the Jeans length is resolved by at least 5 grid cells. 
This refinement strategy enables us to prevent artificial fragmentation \citep{Truelove_97}, 
allows us to efficiently resolve clump formation and evolution and concentrates the resolution
to the central region, but still allows us to trace potential outflows towards several 100\,pc scales. 
A base grid with a cell size of 32\,pc is chosen and with 7 levels of refinement we reach a smallest 
grid size of 0.25\,pc in the dense structures close to the midplane of the disc.
The interactions of the stellar particles within the potential are calculated with a particle-mesh technique.

\subsection{Numerical treatment of star formation}
\label{sec:num_sf}

In this work, we follow the definition of \citet{Krumholz_05} for the 
dimensionless star formation rate per free-fall time $\epsilon_{\mathrm{SF}}$
as the fraction of the mass (above a certain density threshold 
$\rho_{\mathrm{SF}}$) of a grid cell that it converts into stars per 
free-fall time at this density: 
$\epsilon_{\mathrm{SF}} = \dot{M}_{*} \, / \, [M(\ge\rho_{\mathrm{SF}}) 
\, / \, t_{\mathrm{ff}}(\rho_{\mathrm{SF}})]$, where $\dot{M}_{*}$ is the 
star formation rate, $M(\ge\rho_{\mathrm{SF}})$ is the 
mass in the volume where $\rho \ge \rho_{\mathrm{SF}}$ and the 
free-fall time of a sphere is given 
by $t_{\mathrm{ff}} = \sqrt{3\,\pi\,/(32\,G\,\rho)}$ and $\rho$ is the local gas density
within the cell.
For Giant Molecular Clouds (GMCs) $\epsilon_{\mathrm{SF}}$ was observationally found 
to be roughly 0.01 \citep{Zuckerman_74}. \citet{Krumholz_07}
find no evidence for a transition from slow to rapid star formation up to 
densities of $n_{\mathrm{H}}\approx 10^5\,\mathrm{cm}^{-3}$, but the compiled 
observational data is consistent with $\epsilon_{\mathrm{SF}}$ of a few per cent,
independent of gas density. 

To model star formation in the code, we use a modified version
of the {\sc Ramses} standard recipe \citep{Rasera_06}, which we will 
only briefly describe in the following.
Star formation is treated in our implementation 
on a cell-by-cell
basis and takes place, whenever the gas density exceeds the threshold
density for star formation ($\rho > \rho_{\mathrm{SF}}$). Within
this numerical recipe, both, $\rho_{\mathrm{SF}}$ as well as 
$\epsilon_\mathrm{SF}$ are free parameters.
This threshold density is set such to (i) prevent the smooth initial disc from 
forming stars and only allow dense collapsing clumps created by Toomre instability to
form stars and (ii) be below the density threshold at which the artificial 
pressure floor is activated (see Sect.~\ref{sec:numconcepts}), which is resolution dependent. 
Due to the already very high densities of the initial condition and the limited resolution,
this results in a narrow possible range and a star formation 
density threshold of $n_{\mathrm{H}^*}=2\times 10^6\,\mathrm{cm}^{-3}$ was chosen, in 
accordance with \citet{Lupi_15}.
The high 
spatial resolution in these simulations concentrating on galactic nuclei allow 
to resolve the gravitational collapse up to these high densities.
In order to reach a reasonable match with the starburst Kennicutt-Schmitt relation
by \citet{Daddi_10} on global scales (see discussion in Sect.~\ref{sec:KennSchmidt}), we
adjust the star formation efficiency.
The latter criterion links the two SF parameters and is
identical to requiring a global gas depletion time scale in
accordance with observations and a
value of $\epsilon_\mathrm{SF}=0.02\%$ is found.
Those cells fulfilling the criteria form stars according to a Schmidt-like star
formation law: $\dot{\rho}_* = \epsilon_{\mathrm{SF}}\,\rho\,/\,t_{\mathrm{ff}}$, 
where $\rho$ is the local gas density.
This is realised with stellar N-body particles, which
have an integer multiple $N$ of a fixed threshold mass $m_*$, set to 100\,M$_\odot$ 
(see Table~\ref{tab:model_params}). 
Each stellar particle hence corresponds to a stellar population or star cluster.
$N$ is determined following a Poisson
probability distribution function \citep{Rasera_06} 
in order to accurately sample the required star formation rate. 
Whenever a star particle is born, 
we update the fluid and the star particle quantities in a conservative way, the latter
take over the velocity of the gas cell and they are put at the centre of their parent cell. Care is also taken 
that no more than 50\% of the gas is consumed in new-born stars within a single star formation
event. 

It should be noted that such an approach is complementary to models which 
derive $\epsilon_{\mathrm{SF}}$ from first principles, 
like e.~g.~\citet{Krumholz_05}, but
the treatment is in line with the finding by \citet{Krumholz_12} that 
star formation follows a simple volumetric law, depending only on local gas conditions. 

\subsection{Treatment of supernova feedback}
\label{sec:num_sn}

\begin{figure}
  \includegraphics[width=\columnwidth]{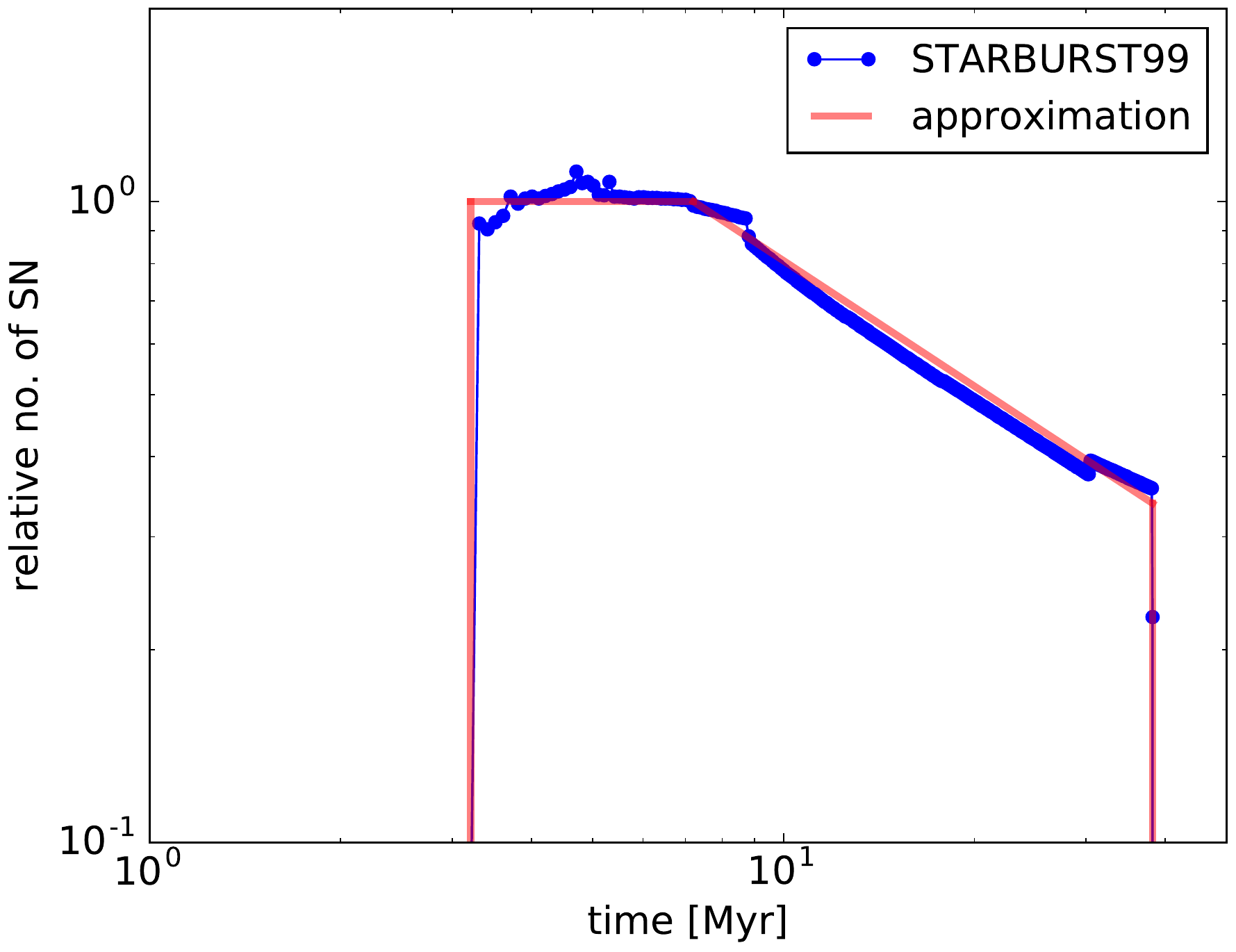}
  \caption{Normalised SN delay time distribution adapted from a {\sc Starburst99} simulation. The red
  line shows the analytical approximation used in the simulations.}
  \label{fig:sn_delaytime_distribution}
\end{figure}

For the supernova feedback, we use a modified version of the {\sc Ramses} implementation
by \citet{Dubois_08}. Star particles are evolved with a
particle-in-cell method and a fraction $\eta_{\mathrm{SN}}$ of their mass will be recycled into the ISM 
during supernova (SN) explosions, whereas the remaining part is locked in long-lived stars. 
Assuming a Salpeter initial mass function (IMF)
leads to a SN yield of roughly 10\%. 
Assuming a typical progenitor star mass of $10\,$M$_\odot$, each star particle
-- which we identify with a stellar population or star cluster -- will cause one 
SN explosion per $m_*$ of star particle mass over time. 
Already during the numerical star formation process described
in Sect.~\ref{sec:num_sf}, we randomly determine a delay time for each prospective SN explosion,
following a SN delay time distribution,
which is the time between the birth of the stellar particle and the detonation of one of its SN.
The delay time distribution is shown in Fig.~\ref{fig:sn_delaytime_distribution}. 
The blue symbols represent the normalised SN rate expected from a coeval stellar population as 
derived from a {\sc Starburst99} \citep{Leitherer_99,Leitherer_14} simulation. The red curve corresponds 
to the analytical approximation used in our simulations. 
SN explosions in this scheme should be thought of as the combined energy and mass input from the stellar wind 
phase and the actual SN explosion. For the most massive stars, the energy input in the stellar wind phase 
can be of similar order or even exceeding the SN explosion itself \citep[e.~g.~][]{Fierlinger_16}. 
Uncertainties in the total ejected SN energies are expected to be at least 
similarly large. 

As soon as a star particle is eligible for an SN explosion, we initiate the following procedure:
\begin{enumerate}
 \item The hydrodynamical state vector is averaged over a given, initial SN bubble radius ($R_{\mathrm{bubble}}$). 
        The latter is set to two cell diameters, in order to be at the same time resolved and smaller
        than typical structures found in the simulations. If the total mass within this radius
        falls short of a mass threshold $m^{\mathrm{SN}}_{\mathrm{threshold}}$, we increase the bubble radius stepwise by 10\%
        up to a maximum radius $R^{\mathrm{max}}_{\mathrm{bubble}}$. This procedure prevents tiny time steps 
        in regions of low gas densities. In these low density regions, SN bubbles would grow to larger sizes on short
        timescale anyway, validating our approach. 
 \item A fraction of 50\% of the total SN energy of $10^{51}$\,erg 
       is injected as thermal energy evenly distributed over 
       the spherical bubble and 50\% is injected in kinetic energy with a linearly increasing radial velocity as a function of 
       distance from the centre of the explosion.
       Additionally, the SN ejecta take over the motion of their parent stellar cluster.
 \item The remaining (hydrodynamical) evolution is followed self-consistently.
\end{enumerate}

\subsection{Gas cooling}
\label{sec:cooling}

As the focus of this work is on the dynamical evolution, we
use a simplistic treatment of the chemical and thermodynamical evolution of the gas.
An adiabatic equation of state with an adiabatic index of $\Gamma = 5/3$ is assumed.
To account for the cooling of the gas, we use one of the
{\sc Ramses} cooling modules, which interpolates the cooling rates within 
pre-computed tables for a fixed metallicity corresponding to solar abundances. 
The latter have been calculated using the CLOUDY photoionisation 
code \citep{Ferland_98}. This results in a comparable effective cooling curve to the one
described in \citet{Dalgarno_72} and \citet{Sutherland_93}. 
A fraction of the gas is expected to be heated by photoelectric heating from 
the forming young stars. In order to save computational time, this is crudely accounted for 
by applying a minimum temperature cut-off at $T=10^4$\,K (see also discussion in Sec.~\ref{sec:discussion}). 

\subsection{Additional numerical concepts}
\label{sec:numconcepts}

If the Jeans length becomes comparable to the grid scale, pressure gradients that stabilise the gas against collapse 
cannot be resolved anymore and artificial fragmentation can occur in the self-gravitating gas. 
To efficiently avoid this, we introduce an 
artificial pressure floor in addition to the temperature threshold mentioned in Sec.~\ref{sec:cooling}
\citep[e.~g.~][]{Machacek_01,Agertz_09,Teyssier_10,Behrendt_15}:

\begin{eqnarray}
P \ge \frac{\rho^2\,G}{\pi\,\gamma}\,N^2\,\Delta x^2
\end{eqnarray}

where $P$ is the thermal pressure, $\rho$ the density of the gas, $G$ the gravitational constant, $\Delta x$ the minimum cell size,
$\gamma=5/3$ the adiabatic index. This pressure floor ensures that the local Jeans length is resolved with at least $N$ grid cells.
We choose $N=4$ in order to fulfil the Truelove criterion \citep{Truelove_97} and thereby avoid artificial fragmentation 
throughout the simulation volume.
The corresponding heating of this numerical treatment, 
however, does not affect the star formation described in Sect.~\ref{sec:num_sf}. 

\section{Results of the simulations}
\label{sec:results}

\subsection{Overall evolution of the simulation}
\label{sec:gasdens}

\begin{figure*}
  \includegraphics[width=\textwidth]{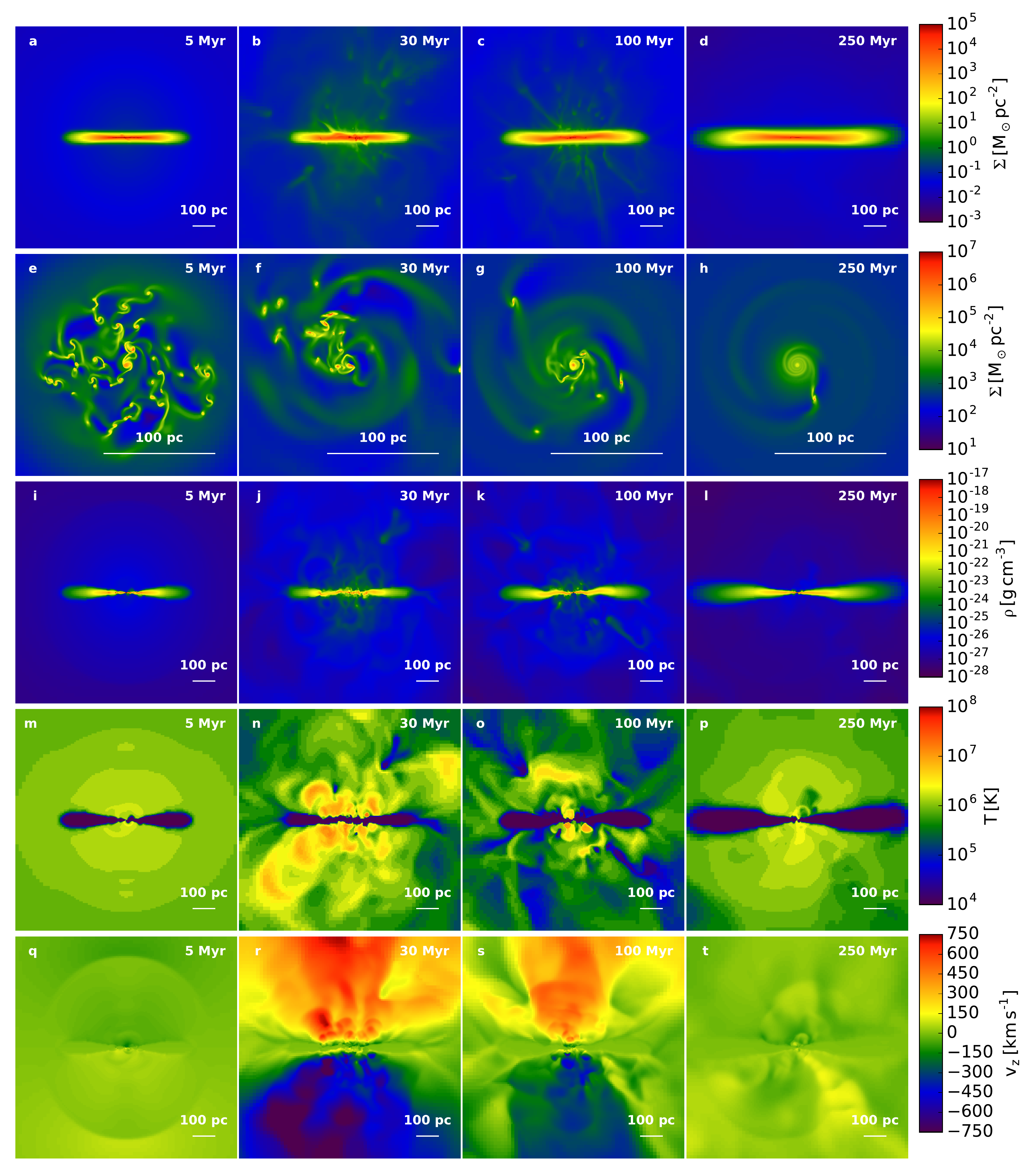}
  \caption{Evolution of the model with time. Shown are density projections along the $x$-axis (upper row), zoomed-in projections
           along the $z$-axis (second row), density slices (third row), temperature slices (fourth row)
           and vertical velocity slices (fifth row) through the $yz$-plane. Mind the different dynamic ranges and scales of the various rows.
           Toomre instability leads to clump formation in the self-gravitating gas disc and the subsequent starburst turns it into a three-component
           structure: (i) the remaining gas disc, (ii) a fountain-like flow in the central region and (iii) a hot, large-scale outflow.
           After the clumps have been dissolved, the SF rate decreases and the disc turns into an almost quiescent state again.
           }
  \label{fig:galnuc_dens_evolution}
\end{figure*}

The time evolution of the gas density distribution is shown in the upper three rows of Fig.~\ref{fig:galnuc_dens_evolution}. 
Density projections along the $x$- and $z$-axis are shown in the first and second row (mind the differences 
in colour and physical scale). The third row shows a cut through the three-dimensional gas density distribution 
along a meridional plane.  
As mentioned in Sect.~\ref{sec:bg_pot_ic_disc}, the initial condition is marginally Toomre unstable in the central 40-50\,parsec
(blue line in Fig.~\ref{fig:galnuc_radial_plots}c).
Small perturbations due to the Cartesian grid allow the growth of unstable modes of this axisymmetric instability, which
leads to the formation of 
a number of concentric ring-like density enhancements. The slightly higher Q-values in the very centre
trigger spiral-like features. During the 
non-linear evolution of the instability, these rings and spiral-like structures break up into a large number of clumps. 
For a detailed description of this evolution we refer to \citet{Behrendt_15}, where Toomre theory is 
studied in great detail with the aim of explaining
the observed properties of gas-rich, clumpy high redshift disc galaxies.
The clumps' early evolution is governed by a complex interplay of various processes: grouping to {\it clump clusters} 
\citep{Behrendt_16}, clump
merging, tidal interactions, (partial) dispersal and gain of mass
by interaction with the diffuse gas component.
After having contracted to reach the gas threshold density, star formation is triggered in their densest, central parts,
leading to depletion of the gas clumps themselves. After the randomly determined delay time, 
the star particles recycle a fraction of their gas to the ISM within SN explosions. These energetic events drive random motions and outflows,
thereby depleting the clumps further. Due to the early strong rise and extremum of the delay time distribution (Fig.~\ref{fig:sn_delaytime_distribution}), 
a three-component flow forms shortly afterwards, especially visible in the edge-on projections of the density (Fig.~\ref{fig:galnuc_dens_evolution}b,c) and 
slices (Fig.~\ref{fig:galnuc_dens_evolution}j,k), as well as the temperature (Fig.~\ref{fig:galnuc_dens_evolution}n,o) 
and $z$-velocity slices (Fig.~\ref{fig:galnuc_dens_evolution}r,s): (i) the remaining thin, high density and cold disc 
(at $10^4$\,K, our minimum temperature) that shows random motions due to clump-clump interactions,  
(ii) a SN-driven fountain-like flow in the central, strongly
star-bursting region where cold dense filaments are ejected into the hot atmosphere  
and partly fall back onto the disc (Fig.~\ref{fig:galnuc_dens_evolution}n,o), stirring additional random motion 
and (iii) a low-density, hot outflow that partly erodes the lifted filaments and escapes the computational domain (Fig.~\ref{fig:galnuc_dens_evolution}n,o and r,s). 
The filamentary structure during the starbursting phase is strengthened, as supernova explosions are more effective in the low, 
inter-filament gas, further enhancing the density contrast \citep{Schartmann_09}.
Overall, this evolution results in a decrease of the clump masses, which is only partly balanced by the merging of clumps. 
Most of the clumps vanish within roughly 200\,Myr, leading to the starvation of the intense star burst (right column of Fig.~\ref{fig:galnuc_dens_evolution}). 
Only a few clumps keep orbiting in the very centre and the system leaves the starbursting regime (see discussion in Sect.~\ref{sec:KennSchmidt} and Fig.\,\ref{fig:kenn_schmidt}). 
As a consequence, the disc returns to an almost quiescent state again with a low number of SN explosions and only a hot and low density outflow remaining, that ceases soon after.

\subsection{Statistical distribution of the gas}

\begin{figure*}
  \includegraphics[width=\columnwidth]{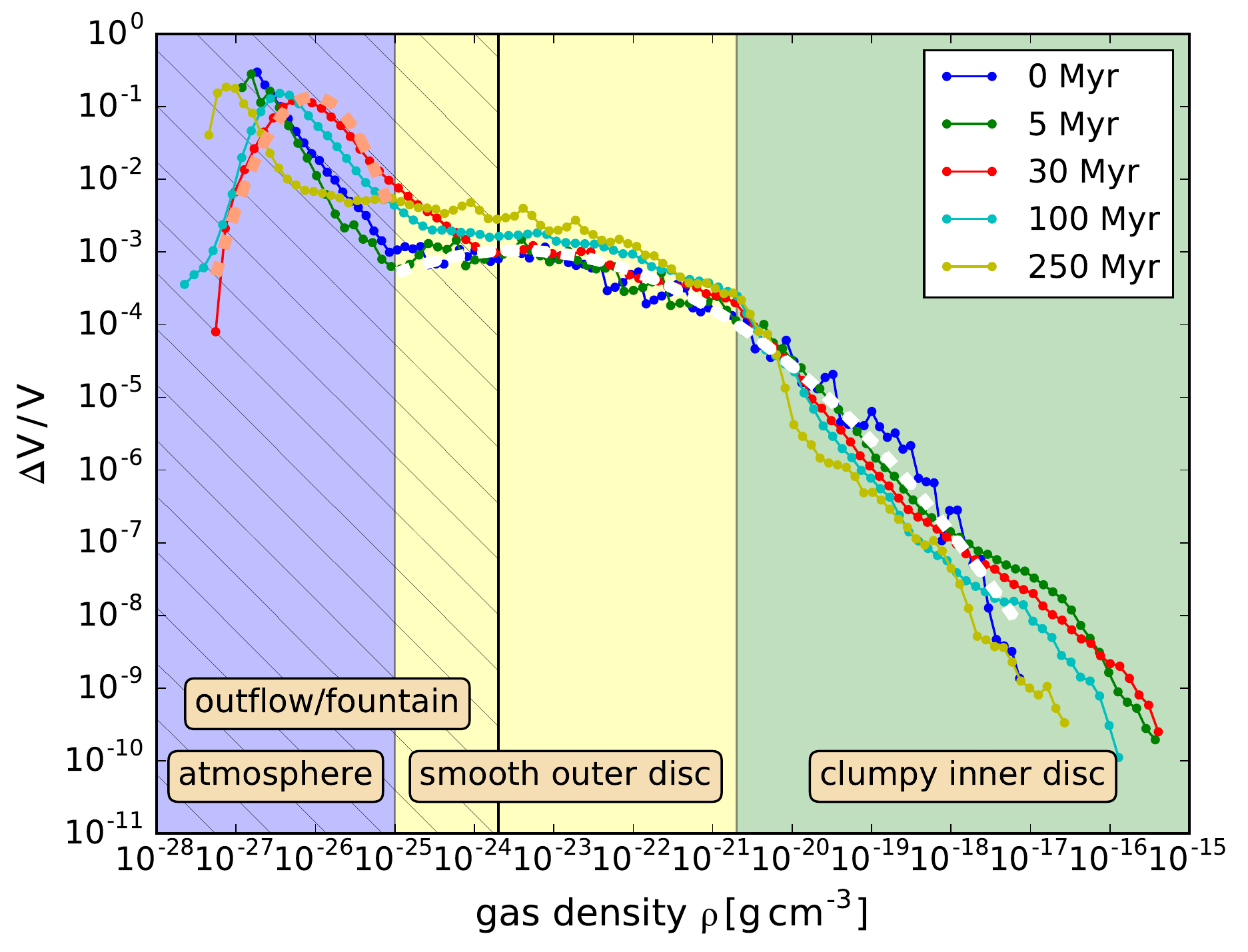}
  \includegraphics[width=\columnwidth]{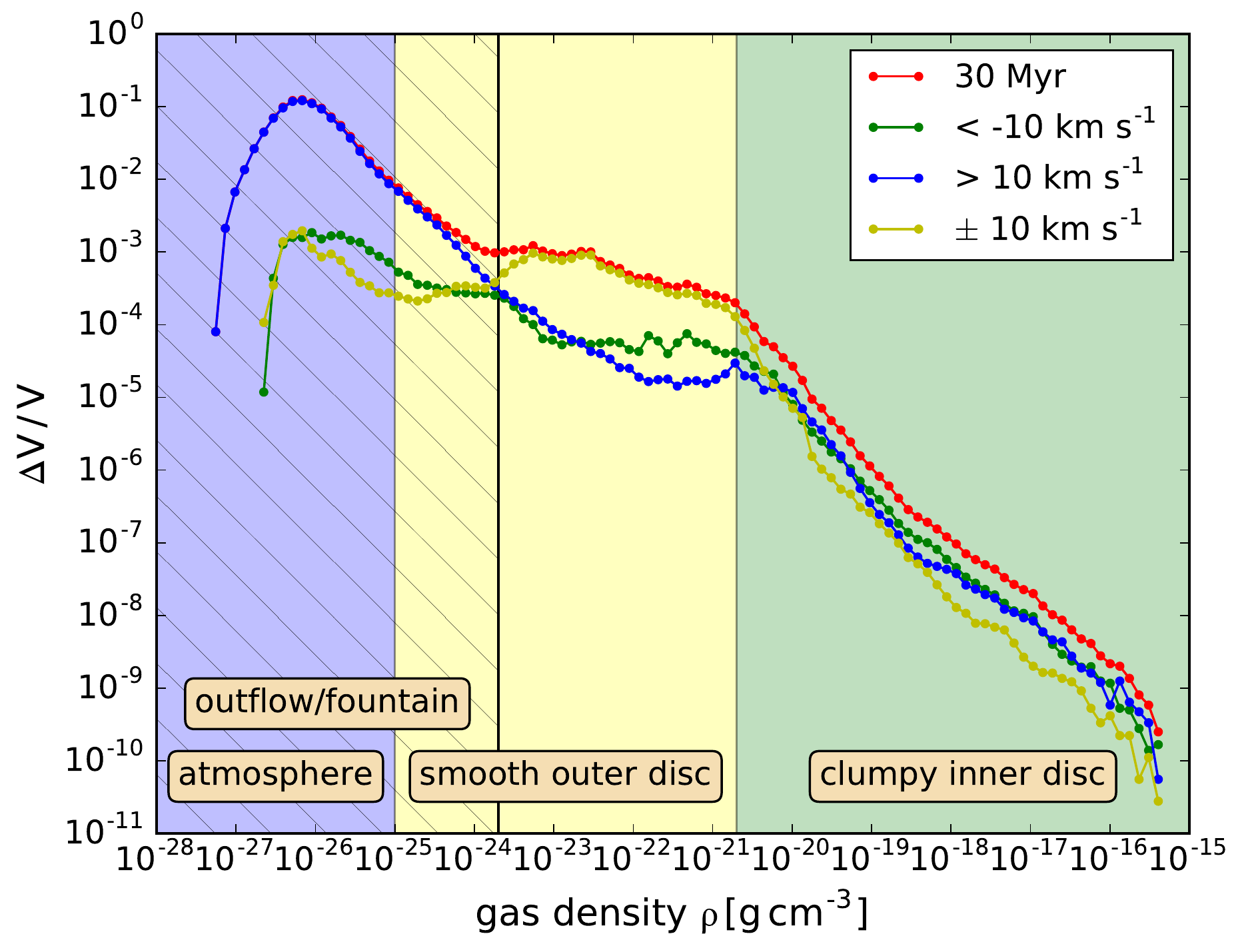}
  \caption{(a) Volume-weighted density probability distribution function (PDF) at several time snapshots as indicated in the legend, marking the initial condition (0\,Myr), the clump formation epoch (5\,Myr), 
           the active outflow / fountain phase (30\,Myr), the less active state (100\,Myr) and the post-starburst, almost quiescent disc (250\,Myr). 
           (b) Density PDF at 30\,Myr split into various radial velocity components
           as indicated in the legend. This permits clear separation of the smooth outer disc from the clumpy, star forming inner disc and the outflow / fountain flow.
           The volume for the plots is limited to a sphere with a radius of 512\,pc
           surrounding the centre in order to remove the edges and outer regions of the cubic simulation box.} 
  \label{fig:dens_PDF}
\end{figure*}

Fig.~\ref{fig:dens_PDF}a shows the time evolution of the volume-weighted density probability distribution functions (PDF) for 
all gas cells within a sphere of 512\,pc radius 
surrounding the centre (disregarding the edges and outer regions of the simulation box) and normalised by the total volume.
The state close to the initial condition is given in dark blue. Two distinct phases can be distinguished: 
the power law at the lowest densities corresponds to the hot atmosphere (marked by the blue background) 
and the smooth initial gas disc (yellow and green background for the initial condition) can be fitted by a log-normal distribution
(white thick dashed line).
The evolution through Toomre instability and the non-linear clump formation leads to a change of the log-normal distribution into a broken-power law distribution
with a knee at around $2\times 10^{-21}$ g cm$^{-3}$, separating the inner, clumpy disc (green background) from the Toomre stable, smooth outer disc 
(yellow background).
Ongoing clump formation and merging extends this power-law to higher and higher densities (green data points), 
until stellar particles are allowed to form.
The power law tail in the high density region of the PDF is expected for self-gravitating gas that can no longer be held up against gravity by turbulence 
\citep[e.~g.~][]{Vazquez_Semadeni_08,Elmegreen_11}.
Around the peak of the starburst (at 30\,Myr, red graph), part of the high density gas phase has been turned into stars and 
energy injection due to the subsequent SN explosions
feeds the fountain flow and hot outflow. This results in a decrease of the volume and mass of the high density gas and a
characteristic bump in the density PDF at lower densities (hatched region), replacing the hot atmosphere in hydrostatic equilibrium.
The latter can be described by a log-normal distribution -- which is characteristic of isothermal, supersonic turbulence \citep[e.~g.~][]{Vazquez_Semadeni_94,Federrath_08} --
plus a higher density power-law. The PDF at this evolutionary stage is split into a radially outflowing, in-flowing and non-moving part (Fig.~\ref{fig:dens_PDF}b). 
This allows clearly separation of the various physical mechanisms and evolutionary states in the simulation:
The remaining smooth disc (or ring) -- unaffected by Toomre instability -- is in radial centrifugal equilibrium and shows almost no in- or outflow motion 
(yellow line in Fig.~\ref{fig:dens_PDF}b) between $5\times 10^{-24}$ and $2\times 10^{-21}$\,g\,cm$^{-3}$. In the inner, clumpy 
part of the disc (green background), the gas is approximately equally distributed between the three kinematic states due to the random motions stirred by clump-clump interactions
following gravitational instability. The bump at low densities can be identified as the wind and fountain flow driven by the starburst, which is dominated 
by random motions. Here, the hot outflowing gas dominates the volume filling fraction, as inflow happens in dense, compressed filaments only.
With decreasing strength of the starburst (cyan line, 100\,Myr; Fig.~\ref{fig:dens_PDF}a), more and more high density clumps vanish and 
the outflow bump moves to lower densities until -- in the post-starburst phase -- the outflow ceases almost completely (yellow line, 250\,Myr) 
towards the end of the simulation.

\subsection{The mass budget, star formation and the stellar distribution}
\label{sec:mass_budget_stars}

\begin{figure}
  \includegraphics[width=\columnwidth]{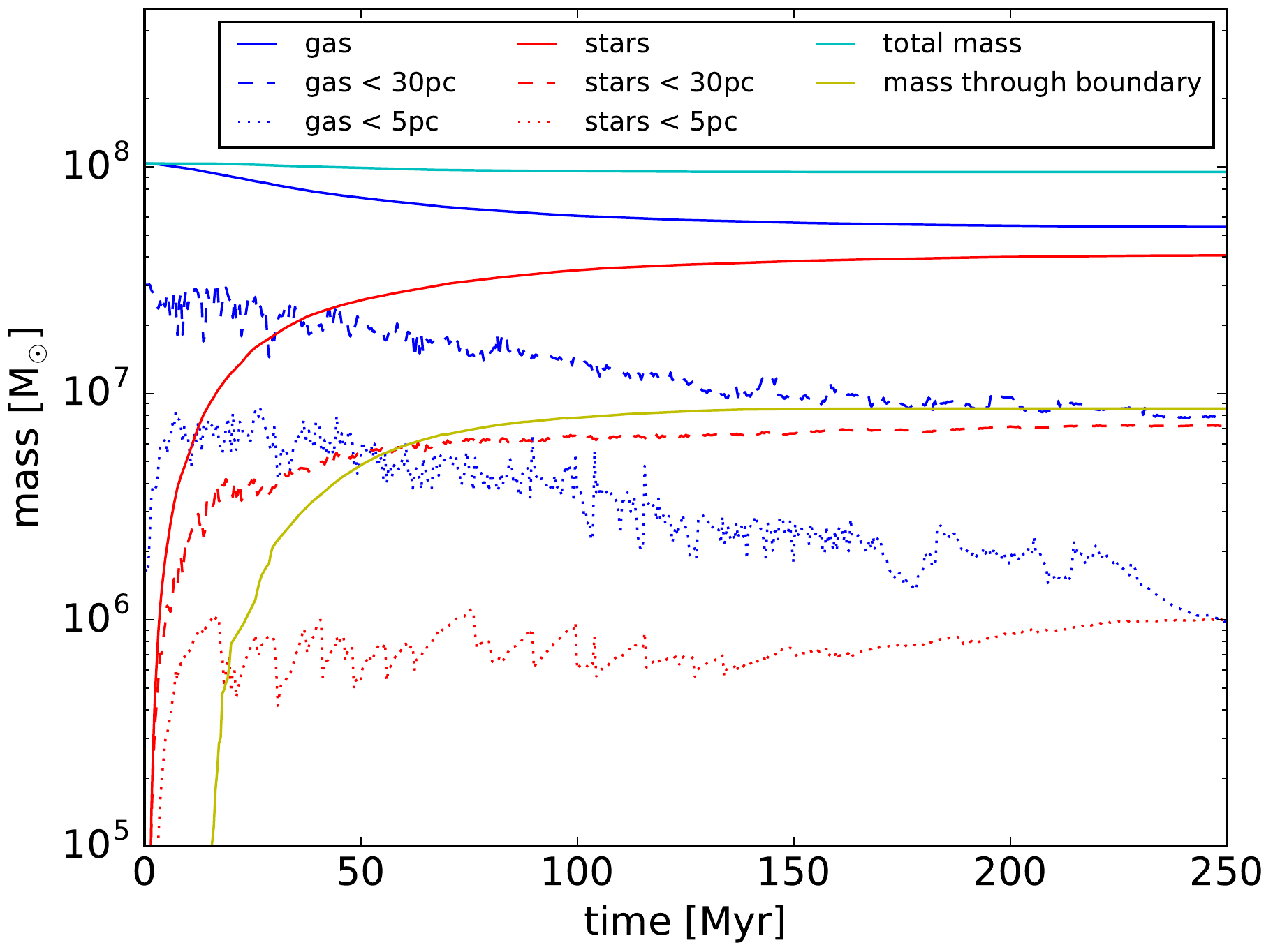}
  \caption{Evolution of the total mass in gas (blue) and stars (red) for the total domain (solid lines) and within a sphere of 30\,pc (dashed lines) and
  5\,pc (dotted lines).
   The yellow line gives the cumulative mass lost through the boundaries of our simulation box. After the active phase of the
   starburst and a stable disc state has been reached, the total mass is still slightly dominated by gas.}
  \label{fig:mass_budget_time}
\end{figure}

Fig.~\ref{fig:mass_budget_time} shows the mass budget of the simulation. 
As soon as the gravitational instability enters the non-linear, clumpy phase, gas is efficiently 
turned into stars. Due to the strongly contracting clumps and high densities reached in the early
evolution of the disc, the resulting SFR 
increases rapidly in the first 10\,Myr, then reaches a maximum of roughly 1\,M$_{\odot}$ yr$^{-1}$, 
followed by an exponential decrease (green stars in Fig.~\ref{fig:mdot_flow_time_selected}). 
The simulation leaves the starburst regime of the Kennicutt-Schmidt plane 
(within the red dotted lines in Fig.~\ref{fig:kenn_schmidt}) after
roughly 100\,Myr. This is given by the dissolution of most of the clumps at that time.
Such a relatively short starburst period is expected given the high gas densities, the short time scale of the fragmentation process
and short gas depletion time in the nuclear region of our galaxy setup. 

\begin{figure*}
  \includegraphics[width=\textwidth]{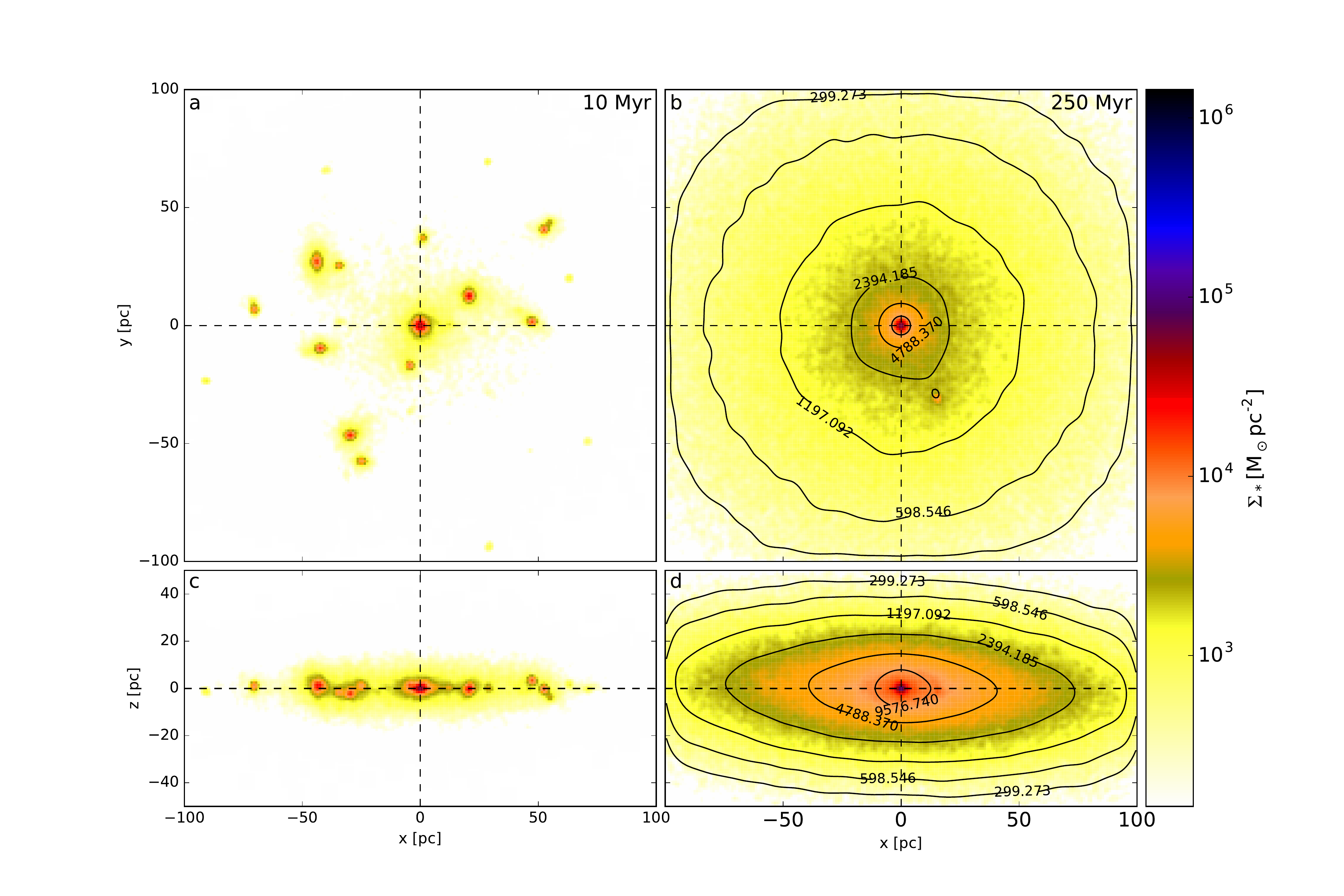}
  \caption{The stellar mass surface density projected along the $z$-axis (a,b) and $y$-axis (c,d) is shown,
  which has been derived on a grid with a bin size of 1\,pc, smoothed with a Gaussian of FWHM=2\,pc. 
  Panels a and c depict the state during the early evolution of the starburst at 10\,Myr and b and d in the almost quiescent, evolved state after 250\,Myr.  
  The black lines correspond to isodensity contours to better visualise the smooth, thick disc structure 
  attained in the late time evolution. To derive the contours, the projected image has been smoothed with a Gaussian of FWHM=10\,pc.
  Whereas the early evolution is characterised by an asymmetric distribution concentrated in a handful of clumps (corresponding to the birth places 
  of the stellar particles), the stars settle into
  a thick disc-like, quasi-equilibrium structure, following relaxation mainly due to clump-clump interactions in the early, clumpy phase of the evolution.}
  \label{fig:galnuc_partdist}
\end{figure*}

Concerning the stellar distribution, we will only discuss the distribution of the stellar particles, that were newly formed in the simulation 
in the following. It should be kept in mind
that they spatially coexist with old stars from the stellar bulge. The latter -- however -- are taken into account as a background potential only.
All of the stars form in dense clumps in our simulations. This is directly visible in the clumpy stellar surface density projected along the 
$z$-axis and $y$-axis during the starburst (see Fig.~\ref{fig:galnuc_partdist}a,c). 
In the following evolution, a thick stellar disc is formed as the result of a relaxation process due to the interaction with the time-dependent
local and global potential (Fig.~\ref{fig:galnuc_partdist}b,d).
Zooming onto and following the evolution of single stellar particles, we find that mergers of their host gas clump with minor gas concentrations
offset the stellar particles from their originally centered position in the host gas clumps. This leads to the gradual heating of the stellar disc and the 
stars start orbiting the merger product.
The strongest interactions occur during mergers of massive clumps, partly leading to a strong time dependence of the local potential and
the ejection of the stellar particle from its host gas clump. Many additional encounters with massive clumps puff up and homogenise the 
stellar distribution to form a thick disc (Fig.~\ref{fig:galnuc_partdist}b,d).  
The decreasing number density of clumps (and clump interactions) with time, 
substantially slows down relaxation, leading to long-lived stellar clumps at late stages of the simulation.
One example is visible in the lower right quadrant of Fig.~\ref{fig:galnuc_partdist}b.

\begin{figure}
  \includegraphics[width=\columnwidth]{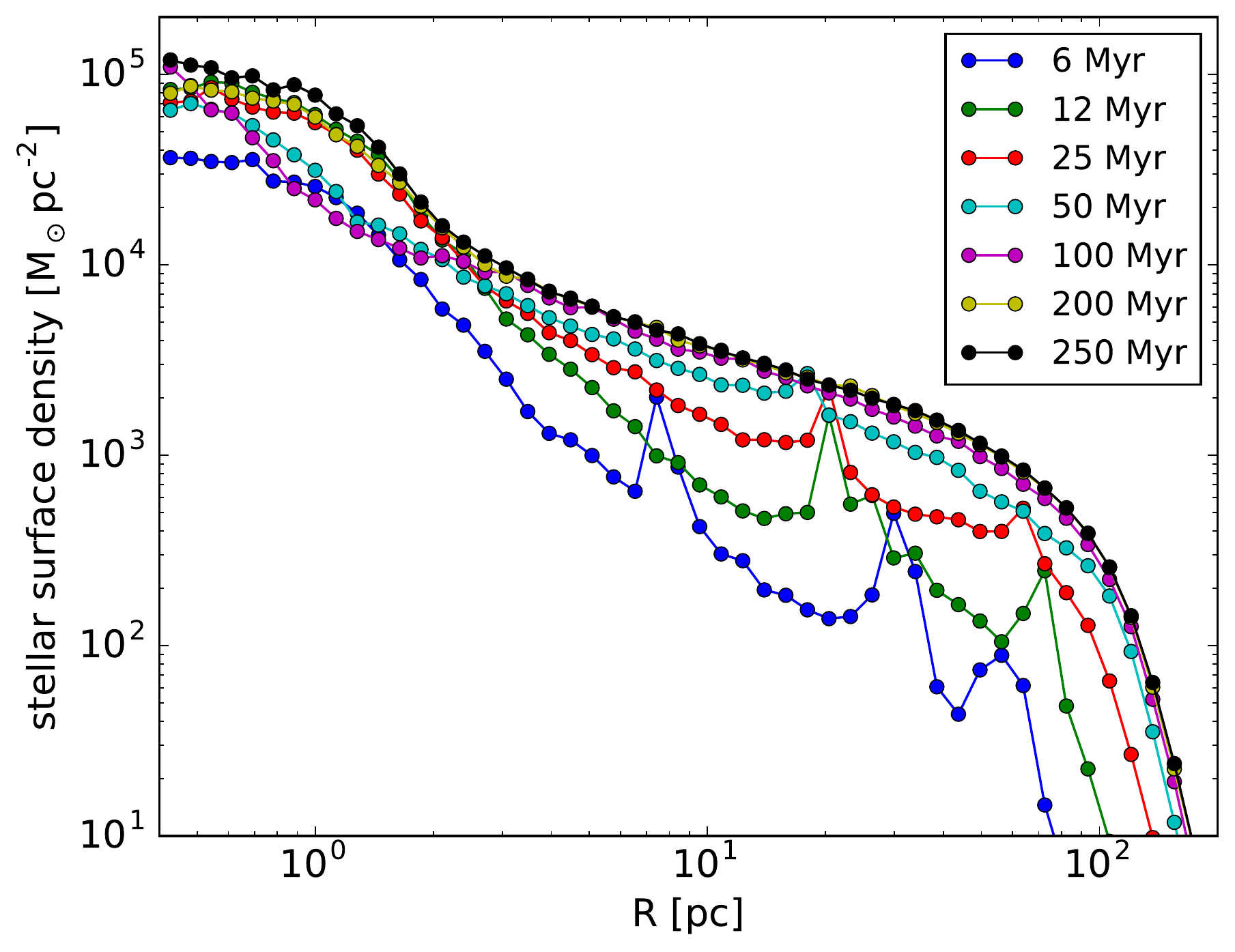}
  \caption{Projected radial stellar surface density profiles. The system relaxes from a distribution
  dominated by a small number of clumps to a converged, homogeneous, thick stellar disc.
  }
  \label{fig:galnuc_particle_radial}
\end{figure}

The relaxation process is best visible in the azimuthally averaged stellar surface density 
distributions (projected along the z-axis) shown in Fig.~\ref{fig:galnuc_particle_radial}, which
also demonstrate that a converged distribution is reached after roughly 100-200\,Myr. 
The process of thick stellar disc formation due to internal processes (clump-clump interactions) observed in our simulations is reminiscent 
of the formation of classical bulges \citep{Noguchi_99,Immeli_04,Elmegreen_08}
and exponential stellar discs \citep{Bournaud_07} through clumpy disc evolution due to violent disc instabilities.
This is found to be a very fast process that lasts only a few rotational periods \citep{Elmegreen_08,Bournaud_09}
consistent with our findings.
Following different dynamics, a less violent process is the drift of the newly formed star clusters with respect to their parent clumps.
This has been observed for the case of the Milky Way Galaxy, where relative drift velocities of around 10\,km s$^{-1}$ have 
been found \citep{Leisawitz_89}. 

Towards the end of the simulation time, the SFR and SN rate have decreased substantially and the system reaches an almost quiescent state again, 
in which the total mass budget is still (slightly) dominated by gas and only very low star formation and outflow rates are maintained. 
Only a small fraction of roughly 10\% of the initial gas mass is lost in the tenuous, hot wind through the outer boundary of our 
simulation domain (yellow line in Fig.~\ref{fig:mass_budget_time}, assuming mass conservation of the code) corresponding to a 
total of roughly $9\times 10^6\,$M$_{\odot}$ within the 250\,Myr computation time.

\subsection{Dynamical evolution of gas and stars}

\begin{figure}
  \includegraphics[width=\columnwidth]{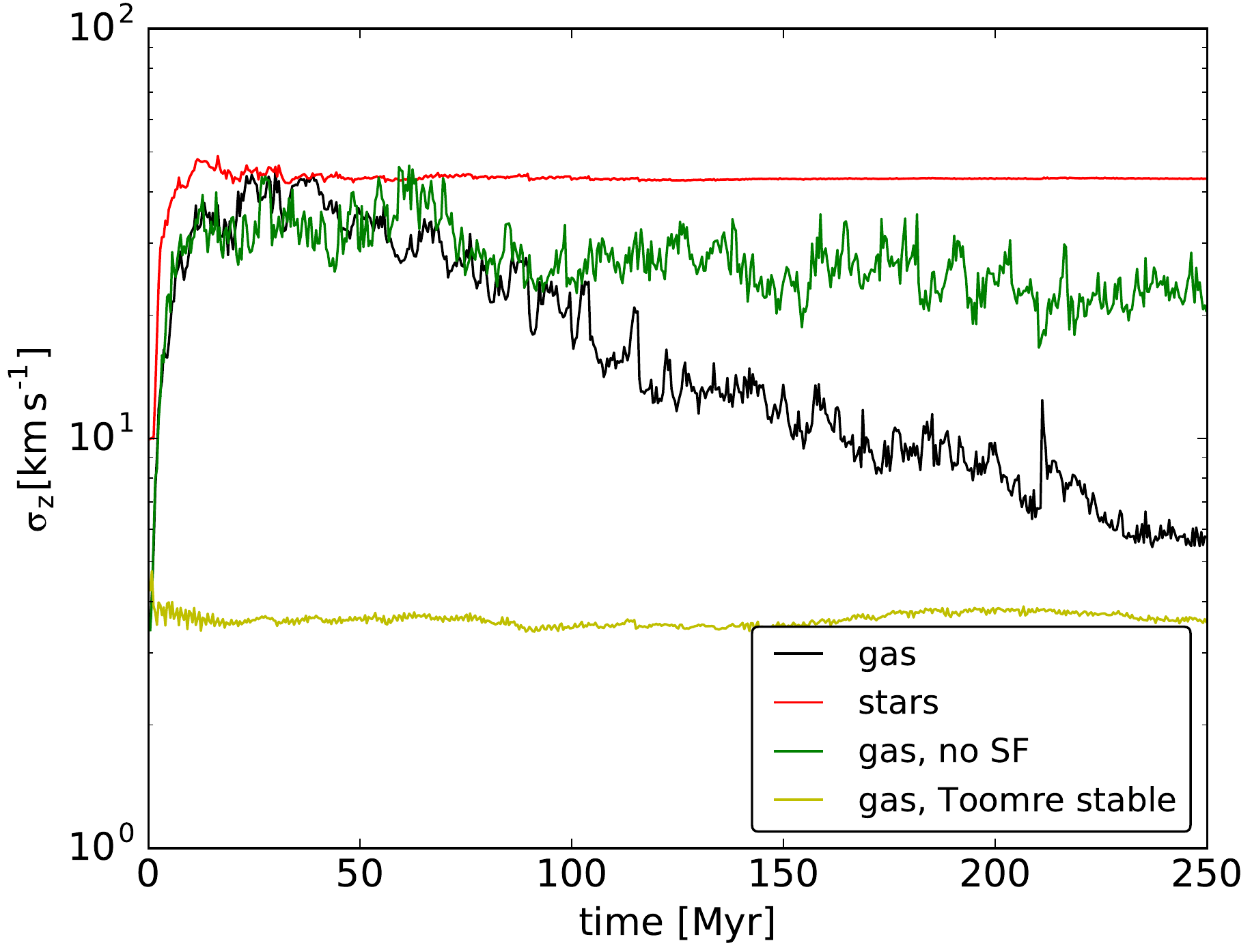}
  \caption{Global mass-weighted gas and stellar velocity dispersion in vertical direction. Clump-clump interactions
           during the early, non-linear stage of violent disc / Toomre instability
           transfer gravitational energy into random motions. The dissipationless
           evolution of the stellar particles enables a boost of the {\it stellar heating} process
           during clump-mergers (red line) compared to the clump-only simulation (green line, without star formation).
           The gas velocity dispersion of the simulation including star formation and feedback approaches 
           the values of a Toomre stable (low mass) control simulation in the late stages
           of its evolution (yellow line).
           }
  \label{fig:galnuc_dispersion_z}
\end{figure}

Fig.~\ref{fig:galnuc_dispersion_z} shows the global mass weighted velocity dispersion of gas and stars in the direction perpendicular to the disc plane. 
The gas component is shown by the black line. The gas dispersion rises steeply during the first 10~Myr. As the gravitational instability 
produces clumps that carry a significant 
fraction of the total disc mass, they interact strongly, leading to a gravitational heating of the disc 
(see discussion in Sect.~\ref{sec:mass_budget_stars}).
To this end, the simulation without star formation (green line) shows a similar 
rising signature. Whereas the simulation without SF remains at roughly the same level (given by the relaxation of the gas clumps in the global
potential and sustained by random motions of the clumpy medium with a low volume filling factor), 
the simulation with SF and SN feedback shows a decreasing trend of the gas dispersion. 
This is caused by the slow dissolution of the gas clumps caused by the ongoing star formation and feedback and the dissipative interaction
with other clumps and the remaining smooth disc. The SN-driven fountain flow and outflow
mostly affects the lower density gas and hence does not contribute substantially to the mass weighted velocity dispersion.
With the dissolution of most of 
the clumps and the small remaining SN rate, the gas distribution settles into a thin disc configuration, resulting in a small
value of the vertical gas velocity dispersion, slowly approaching the control run of a (low mass) Toomre stable disc (yellow line). 

The stellar, vertical velocity dispersion (red line) shows a very similar behaviour to the gas dispersion in the no-SF case. 
Stars form mainly in the central regions of gas clumps within the equatorial plane of the disc, where the gas density is highest. 
Scattering events of stars with gas clumps -- especially during clump merger events -- lead to enhanced gravitational heating 
compared to the gas distribution. Finally, the stellar distribution relaxes in the global potential 
within roughly 30-40~Myr, leading to a constant velocity dispersion with time of roughly 40~km s$^{-1}$. This is slightly higher than the 
gas velocity dispersion in simulation no-SF in the second half of the simulation. 
We attribute the difference to the collisionless nature of the stellar particles, whereas the gaseous clumps dissipate energy in clump collisions
and when moving through the smooth, ambient inter-clump gas. 

\subsection{Supernova feedback and starburst evolution}

\begin{figure}
  \includegraphics[width=\columnwidth]{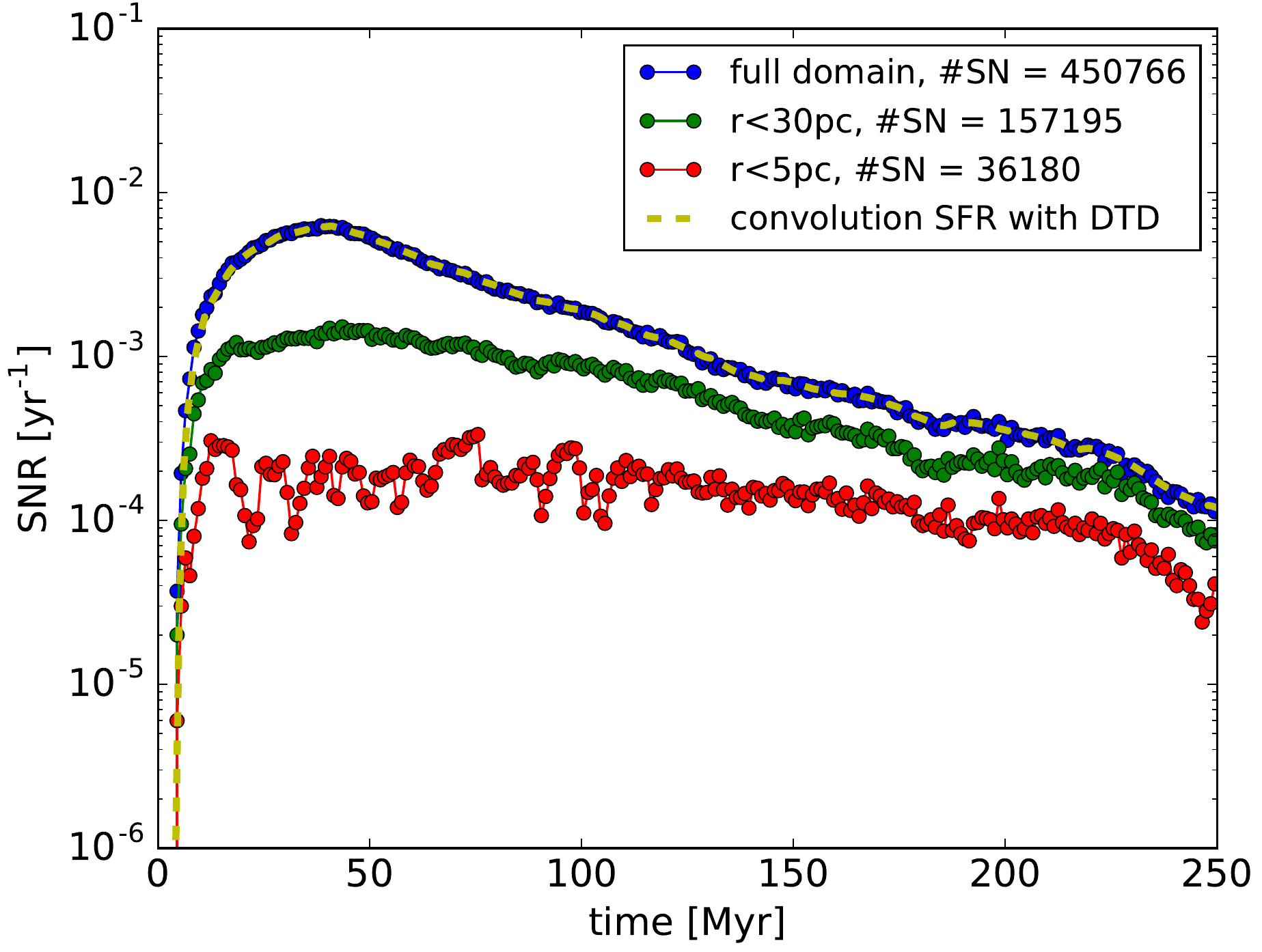}
  \caption{SN rates within the full domain and spheres of several radii as indicated in the legend.
           The global SN rate roughly follows the yellow dashed line, which depicts the convolution 
           of the global star formation rate (SFR; Fig.~\ref{fig:mdot_flow_time_selected}, green stars)
           and the normalised supernova delay time distribution (DTD; Fig.~\ref{fig:sn_delaytime_distribution}).}
  \label{fig:snr_bin_time}
\end{figure}

The global supernova rate is shown with blue symbols in Fig.~\ref{fig:snr_bin_time}. 
It can be roughly understood as the convolution of the global SFR 
(Fig.~\ref{fig:mdot_flow_time_selected}, green symbols) 
with the normalised SN delay time 
distribution (DTD; Fig.~\ref{fig:sn_delaytime_distribution}), which is shown as the 
yellow dashed line
in Fig.~\ref{fig:snr_bin_time}. To this end, it shows a steep rise. The maximum is 
reached at around 30-40\,Myr and -- due to the convolution with the DTD -- is delayed
with respect to the one of the star formation history. 
The supernova rate (SNR) then follows an exponential decline until most of the dense enough gas
has been used up for star formation and most of the massive stars have exploded as SN.
Fig.~\ref{fig:snr_bin_time} indicates that the supernovae in our simulation are strongly 
concentrated to the central region, which we show by restricting
the calculated SNR to within spheres with certain radii (green and red circles). 
The concentration to the central region is expected due to the exponential profile 
of the initial gas distribution that only leads to a Toomre-unstable, central region 
of a few tens of parsecs (Fig.~\ref{fig:galnuc_radial_plots}c). 
Viscous radial gas inflow during the evolution of the simulation 
further enhances this central concentration.
The resulting supernova distribution follows the stellar
distribution which evolves towards a double power-law radial profile 
with a central stellar density concentration
(Fig.~\ref{fig:galnuc_particle_radial}).
The more we concentrate towards the nuclear region, the shallower the distribution  
and the stronger the fluctuations with time. 

The fast relaxation of the stellar particles by clump-clump interactions allows many
of them to leave their host clump and enables the formation of a stellar distribution with 
a moderately large scale height in vertical direction (Fig.~\ref{fig:galnuc_partdist}). The subsequent 
SN explosions can then take 
place in a lower density environment, enabling higher efficiency to stir random 
motions and outflows. In contrast, most of the energy introduced due to SN in high
density environments will be lost due to strong cooling, as the optically thin cooling
applied in our simulations scales proportionally to the square of the gas density.
Most of the early SN explosions will be located in the dense clumps and hence have only
a small effect on the overall gas dynamics, whereas the progenitors of most of the 
later supernovae have already left their parent gas clump and are able to drive substantial
random motions and outflows without major loss of energy via cooling radiation.
The stellar relaxation and migration process hence changes the energy input mechanism 
substantially (see discussion in Sect.~\ref{sec:outflow} and \ref{sec:comp_sim}).

\subsection{Starburst-driven outflows and fountain flows}
\label{sec:outflow}

\begin{figure*}
  \includegraphics[width=0.75\textwidth]{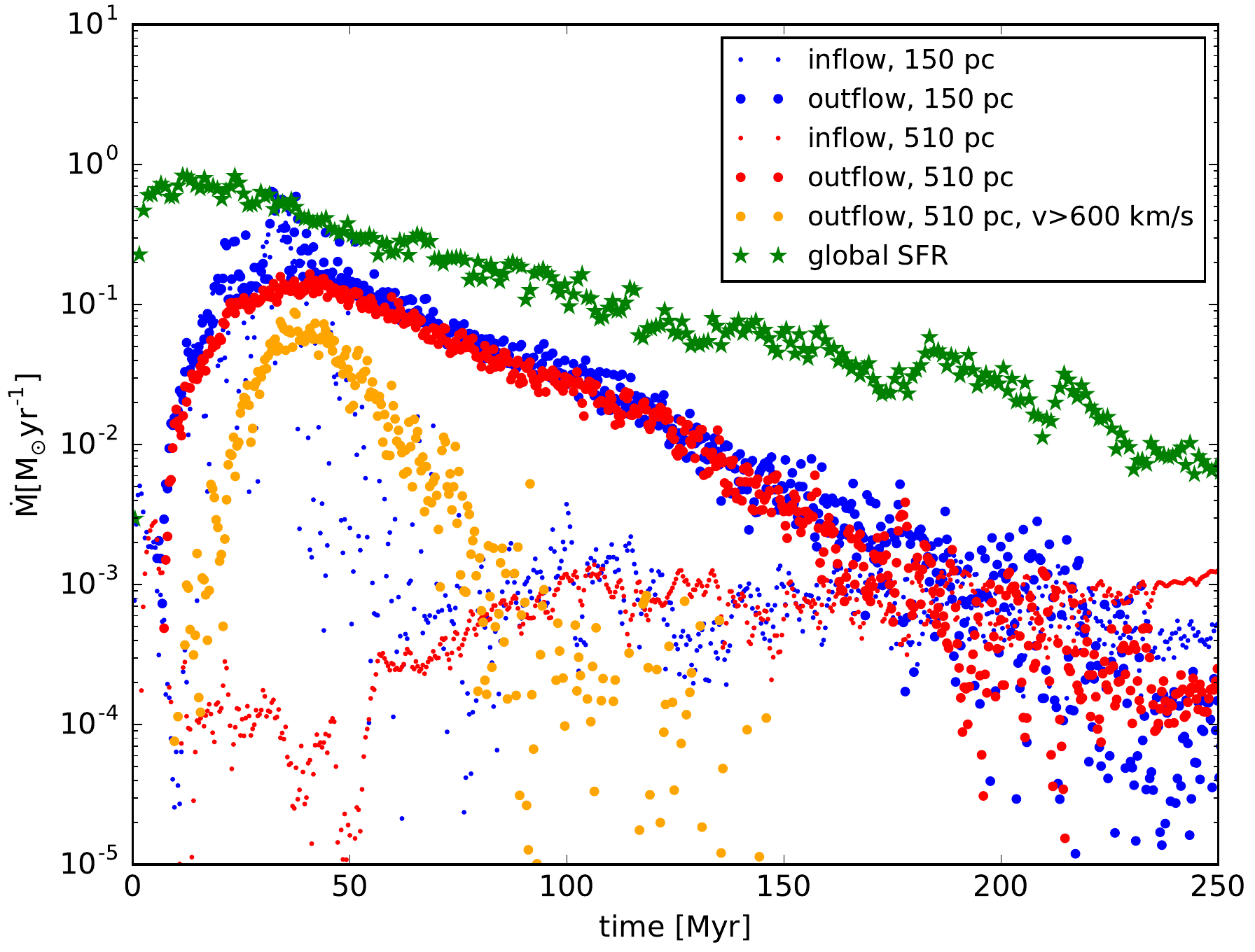}
  \caption{Global star formation rate and inflow / outflow through spherical shells at
  distances of 150\,pc and 510\,pc from the centre for the full evolution time of the simulation.
  The large-scale outflow (red thick circles) is directly
  correlated with the evolution of the SN rate (Fig.~\ref{fig:snr_bin_time}) during the most
  active phase of the starburst.
  The early evolution shows a fountain-like flow at small distances from the centre
  with similar in- and outflow rates (blue thin and thick circles) that turns into a strong
  outflow. Only during the peak of the starburst, a significant fraction of the outflow
  reaches escape velocity from the galaxy (orange symbols, for an assumed escape velocity of 600\,km\,s$^{-1}$).}
  \label{fig:mdot_flow_time_selected}
\end{figure*}

During the peak of the starburst, the strong energy input in supernova explosions drives both a low
gas density outflow as well as a higher density fountain like flow (see Fig.~\ref{fig:galnuc_dens_evolution}).
In order to quantify the 
two phenomena, we measure instantaneous gas in- and outflow rates from the  
simulation snapshots through two spherical shells at a radial distance from the centre of 
150 and 510\,pc with a thickness of $\Delta r=20$\,pc:

\begin{equation}
\dot{M} = \frac{1}{\Delta r}\,\sum_\mathrm{shell} m_\mathrm{i}\,v_\mathrm{i}
\end{equation}
where $m_\mathrm{i}$ and $v_\mathrm{i}$ are the mass and velocity of each cell within the spherical shell.

Fig.~\ref{fig:mdot_flow_time_selected} shows the mass transfer rates calculated in this way as a function of time.
This procedure allows us to quantify inflow (thin dots; radial velocity smaller than -10\,km s$^{-1}$) as well as outflow 
(thick dots; radial velocity larger than 10\,km s$^{-1}$) motion.  
The red symbols depict the measurements
for the outer shell located at 510\,pc distance from the centre.
The outflow through this shell lags behind the star formation rate (green stars) and 
reaches its maximum after roughly 30-40\,Myr following a shallower increase compared to the SFR. 
This delayed maximum of the outflow can again be explained by the DTD of the SNe and closely
follows the SN rate with time
with a slightly shallower increase towards the maximum, which we attribute to 
a combination of effects:
(i) The efficiency of the energy input increases from the early SN explosions within 
dense clumps to SN explosions in the inter-clump medium due to stellar migration 
that heats the particle distribution by interaction with gas clumps. 
(ii) The high density fountain flow at low latitudes suppresses an outflow due to its relatively high gas 
column densities (see discussion in Sect.~\ref{sec:obs_prop}). 

As the large-scale wind is driven from the central (initially Toomre-unstable) 50-100\,pc region, the morphology resembles a 
cylindrical outflow during the most active phase (Fig.~\ref{fig:galnuc_dens_evolution}r,s). 
This is in contrast to radiatively-driven outflows from the central accretion 
disc which result in a more conical wind shape \citep[e.g.][]{Wada_12,Wada_16}.
Only around the peak of the starburst, a significant fraction of the outflow reaches escape velocity from the galaxy. This is 
shown by the orange symbols for an assumed escape velocity of 600\,km\,s$^{-1}$.
Looking at the shell close to the disc (blue symbols), the thin and thick symbols are close to one another for 
the first 40\,Myr. 
This is mainly due to the fountain-like flow of the denser filaments that reaches beyond the
shell position. These are part of the cold filaments visible in 
Fig.~\ref{fig:galnuc_dens_evolution} around that time.
After roughly 50\,Myr -- beyond the peak of the most active starburst -- a steady
outflow is formed with almost no back flowing filaments, even at the location of 
the inner shell. In- and outflow rates only become of similar strength after the starbursting phase 
at around 175\,Myr.

\section{Comparison to observations}
\label{sec:comp_obs}

\subsection{Evolution in the Kennicutt-Schmidt diagram}
\label{sec:KennSchmidt}

\begin{figure}
  \includegraphics[width=\columnwidth]{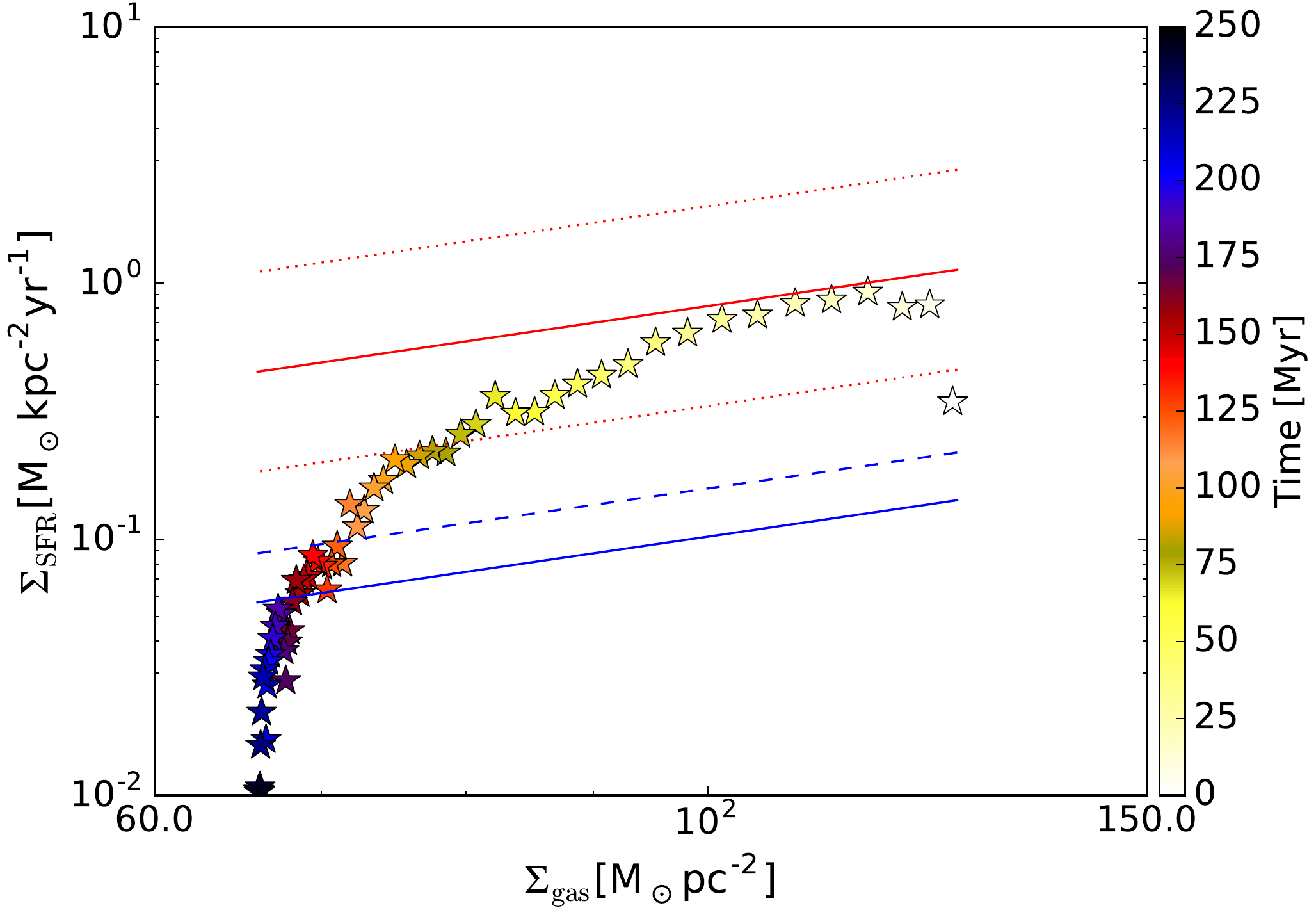}
  \caption{Evolution of the simulation in the Kennicutt-Schmidt diagram, colour-coded with the 
    simulation time, averaged over 4\,Myr. 
    The quiescent KS relation derived by \citet{Daddi_10} is shown as the blue solid line 
    whereas the red one refers to their {\it starburst} KS relation. The red dotted lines roughly give 
    the observed scatter around the relation, the lower one roughly corresponding to the separation
    between bursting and quiescent star formation.
    The original KS relation \citep{Kennicutt_98} is shown as the blue dashed line. Our simulation 
    approaches the starburst KS relation from high gas surface densities, then evolves along the relation 
    (by construction) and the depletion of the clumps leads to a deviation from the relation with time, 
    before the SFR drops steeply after most of the clumps have dissolved.}
  \label{fig:kenn_schmidt}
\end{figure}

Fig.~\ref{fig:kenn_schmidt} shows the evolution of the simulation
in the plane given by the star formation rate surface density and 
the total gas surface density. To derive the data points, averages have
been taken within a fixed (cylindrical) radius of 512\,pc.
This size ensures that all stars are included in 
the averaging process and the low density atmosphere within the simulation box is removed. 
Due to the short time scale of the 
structure and star formation process, both, the gas as well as the SFR surface 
densities increase rapidly until they approach the starburst Kennicutt-Schmidt (KS) 
relation \citep{Daddi_10}. The simulation starts at very high gas surface densities
and first roughly follows the relation towards lower SFR surface 
density. This is by construction and used to calibrate our 
choice of the star formation efficiency. Then, 
the depletion of gas within the clumps due to star formation and stellar feedback
processes results in a decrease of the average gas surface density. 
With the dissolution and merging of more and more clumps, 
the simulation starts deviating from the observed relation and leaves the 
starburst regime (within the red dotted lines) after roughly 75-100\,Myr. 
After another approximately 50\,Myr, 
most of the clumps have dissolved and only a handful of them remain in 
the central region.
Observationally, gas depletion times have been found to be shorter in galactic 
centres compared to the rest of the galaxy \citep[e.~g.~][]{Leroy_13}.
In the framework of our model, the duration of the starburst is a consequence 
of the marginally unstable initial condition, in which clumps are only formed in the 
central roughly 100\,pc region of the gas disc and the short gas depletion time scale 
is equivalent to the starburst Kennicutt-Schmidt relation.
Even shorter, Eddington-limited starbursts of the order of 10\,Myr have been inferred 
from observations of circumnuclear discs of a sample of nearby Seyfert galaxies 
\citep{Davies_07}.

\subsection{Comparison to nuclei of nearby (active) galaxies}
\label{sec:davies_hicks_comparison}

The gas mass within the central 30\,pc region (corresponding to the scale length of 
the initial, exponential gas disc) is shown in Fig.~\ref{fig:mass_budget_time} by the blue dashed line. It decreases slightly with 
time, reaching a value of around $10^7\,$M$_{\odot}$, in good correspondence to the observed Seyfert galaxy sample in 
\citet{Hicks_09}. For the same sample, measurements of the velocity dispersions of the warm molecular hydrogen ($\approx$2000\,K) via
the 2.12\,$\mu$m 1-0 S(1) line result in values of 50-100\,km\,s$^{-1}$. For the cold gas phase represented by 
the 3mm HCN(1-0) line, lower velocity dispersions of 20-40\,km\,s$^{-1}$ are derived \citep{Sani_12}. 
\citet{Lin_16} probe the velocity dispersion of the dense gas ($n_{\mathrm{H_2}} \approx 10^{4-5}\,\mathrm{cm}^{-3}$), 
resulting in a median velocity dispersion of $35$\,km\,s$^{-1}$. 
These values for the cold, dense gas are compatible to our mass-weighted velocity dispersions shown in 
Fig.~\ref{fig:galnuc_dispersion_z}, which probe the high density component of the multi-phase gas in the simulations.

Using Keck/OSIRIS near-infrared (NIR) spectra, \citet{Durre_14} investigate the circumnuclear disc in NGC\,2110
\citep[e.~g.~][]{Veron_06,Rosario_10,Storchi-Bergmann_99}
and find a total, nuclear star formation rate of $0.3\,$M$_{\odot}\,\mathrm{yr}^{-1}$. 
This SFR is dominated by four massive and young star clusters that are embedded into a rotating nuclear disc 
of shocked gas in the inner 100\,pc surrounding the active nucleus \citep[see Fig.~2a,][]{Durre_14}. 
The shocked intercluster medium is thought to be excited by strong outflows that do not appear to originate from the AGN, but
rather are localized to the clusters. This is reminiscent of the morphology of the stellar surface density in our simulations
in the phase 
following Toomre instability and clump merging, that shows a handful of clumps orbiting in the nuclear potential in our
simulations (e.~g.~Fig.~\ref{fig:galnuc_partdist}a). These observed clusters could, therefore, be a sign of a 
recent event that triggered gravitational instability and the observed starburst and might have significant impact on the 
(ongoing and future) central activity \citep{Davies_07}.

The stellar distribution as well as kinematics of the central 10-150\,pc in active and inactive galactic nuclei within
the LLAMA (Luminous Local AGN with Matched Analogues, \citealp{Davies_15}) sample has been analysed by M.~-Y.~Lin et al.~(2017, submitted).
After subtraction of the underlying stellar bulge distribution, a stellar light excess is found in most of the sources,
which amounts to a few percent of the stellar mass of the underlying bulge within the central 3\,arcseconds. 
This excess emission is found to be consistent with rotating stellar nuclear 
discs, which follow a size-luminosity relation in which the size of the stellar system is roughly proportional to the 
square root of the stellar luminosity. For the final snapshot of our simulation at 250\,Myr, we find
that 99\% of the stellar mass is inside a radius of 173\,pc. Assuming a mass-to-light ratio
of 1.5 for this relatively young stellar population (see Fig.~4c in \citealp{Davies_07}) results in a total luminosity 
of $3 \times 10^7\,$L$_\odot$. This places the final state of our simulation very close to the observed relation. During the early evolution,
the simulation is located above the relation (but already within the observed scatter after a few Myrs), 
and approaches it with time.

On top of the large scale stellar disc, a second component is found in the central few parsec regime
(Fig.~\ref{fig:galnuc_particle_radial}). 
The latter is characterised by fitting a S\'ersic profile up to a distance of 2\,pc. We find a S\'ersic index 
of 0.6, an effective radius of 1.2\,pc and a total mass of roughly 
$8\times 10^5$\,M$_\odot$ after 250\,Myr of evolution.
This is reminiscent of nuclear star clusters (NSC) that are frequently observed in the centres 
of galaxies of all Hubble types. Their observed dynamical masses are in the range of 
$10^4$ to $10^8$\,M$_\odot$ with effective radii between $0.1-100$\,pc \citep[e.~g.~][]{Balcells_07,Kormendy_10,Georgiev_16}.
Our simulated NSC is within the scatter of the mass-size relation the latter authors find, 
located rather at the lower mass and size end of the distribution.  
The star formation histories are observationally found to be rather long and quite complex
and seem to require multiple epochs of recurrent star formation \citep{Neumayer_12}. 
Together with the finding of NSC rotation as a whole \citep{Seth_08,Seth_10}, these
observed characteristics nicely match the stellar system found in our simulation, 
which might correspond to the first star formation epoch. Subsequent disc instabilities
(following additional mass inflow from the galaxy)
might lead to a growth in mass and size. 
However, it should be kept in mind that at these distances from the centre we are 
within the smoothing scale of the central SMBH potential, which makes further 
investigations necessary.

With the advent of the ALMA observatory, more and more high spatial and spectral resolution data
on nearby AGN becomes available. A particularly interesting case is the Circinus galaxy, the nearest
Seyfert~2 galaxy. The molecular gas in its central 1~kpc region has been studied
in great detail in CO~(1-0) line emission by \citet{Zschaechner_17}. 
They find evidence for a molecular component in the well-established ionised gas outflow with an 
outflow rate of 0.35 to 12.3\,M$_{\odot}$, which is comparable to its (global) star formation rate of 4.7\,M$_{\odot}$\,yr$^{-1}$
\citep{For_12}. This is regarded as an indication that the outflow regulates the SF. A velocity dispersion of 30-50\,km\,s$^{-1}$ has 
been found in the central roughly 5~arcsec and 30-40\,km\,s$^{-1}$ in the outflow.
A total molecular mass of the disc within 1.5\,kpc of 2.9$\times 10^8$\,M$_{\odot}$ has been inferred. 
All of these numbers -- as well as the hint for a filamentary structure in the wind -- are roughly compatible to what we find in 
our simulations during the actively starbursting episode.
However, in our simulations, the star formation regulates the outflow and not vice-versa and the outflow rates 
are close to the lower limit of the observationally derived outflow rate, allowing for additional outflow driven by the active
nucleus.

IC\,630 is a nearby early-type galaxy, classified as a radio galaxy \citep{Brown_11}. With the help of VLT-SINFONI and Gemini North-NIFS adaptive 
optics observations, 
\citet{Durre_17} show that the excitation of the circumnuclear gas within a few 100\,pc of the 
centre can mostly be explained by star formation. The measured SFR (1-2\,M$_{\odot}$\,yr$^{-1}$), SNR ($4\times10^{-3}$\,yr$^{-1}$), the gas outflow rate 
(0.18\,M$_{\odot}$\,yr$^{-1}$), total gas masses (a few times $10^7$\,M$_{\odot}$) as well as the line of sight 
velocity channel maps of the gas all are in line (within a factor of a few) with our simulated system during the peak of the starbursting phase, 
seen face-on. Especially the inferred young starburst age of roughly 6\,Myr (derived from the equivalent width of the Brackett-$\gamma$ line) and 
the observed cylindrical rather than conical shape of the outflow points into a starburst-, rather 
than AGN-driven event. Also the clumpy structure in the observed NIR J, H and K band maps as well as the hint of a spiral 
structure is in good agreement with the structural evolution found from our violent disc instability scenario.

Overall, we find good agreement with available observations, indicating that the physical scenario investigated in this paper could be at work
in several nearby (active) galactic nuclei.

\section{Discussion}
\label{sec:discussion}

\subsection{Obscuring properties}
\label{sec:obs_prop}

\begin{figure}
  \includegraphics[width=\columnwidth]{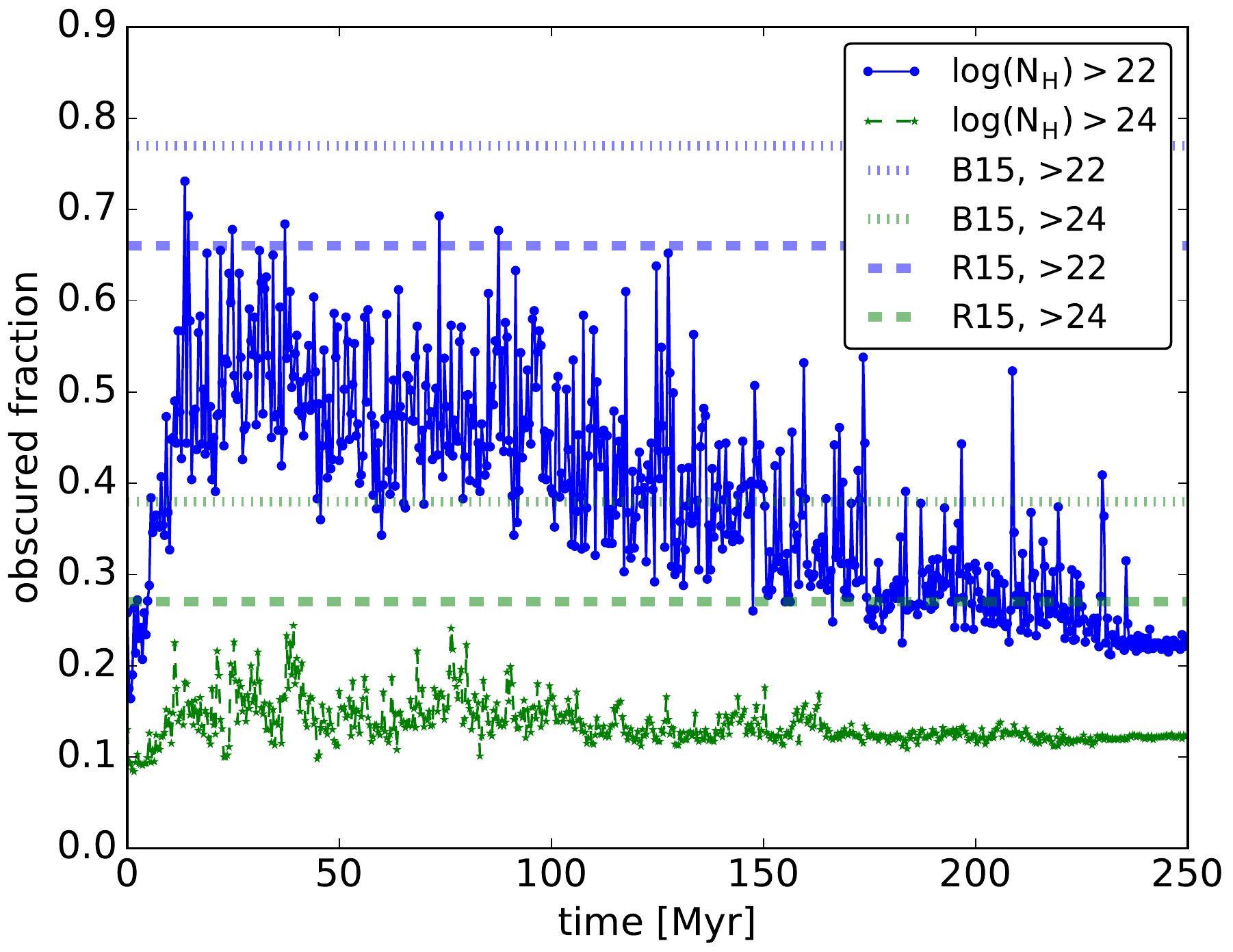}
  \caption{Obscured fraction as a function of time for log(N$_\mathrm{H}$[cm$^{-2}$])>22 
  (circles and blue solid lines) and log(N$_\mathrm{H}$[cm$^{-2}$])>24 
  (stars and green dashed lines). The inner region has been cut out up to a radius of 5\,pc. 
  Shown as well are the values derived from observations by \citet{Buchner_15} (B15) and \citet{Ricci_15} (R15).
  This shows that a nuclear starburst on tens to hundreds of parsec scale is able to significantly 
  contribute to the obscured fraction and leads to a time varying additional obscuration.
  }
  \label{fig:obscured_fraction}
\end{figure}

Fig.~\ref{fig:obscured_fraction} shows the obscured fraction as a function
of time for the gas distribution of our simulation. For each data point, 
we calculate 1000 rays from the centre to determine
the hydrogen column density. Shown is the resulting obscured fraction
defined as the fraction of these rays having a hydrogen column density in excess of 
$10^{22}\,\mathrm{cm}^{-2}$ (blue circles) and $10^{24}\,\mathrm{cm}^{-2}$ (green stars). 
Due to resolution limitations we are currently not able to resolve AGN tori or 
nuclear maser emitting discs. In the nearby
active galaxies we are concentrating on, the latter are  
expected to have sizes in the sub-pc to pc 
regime \citep[e.~g.~][]{Greenhill_96,Greenhill_03,Burtscher_13,Tristram_14,Garcia_Burillo_16}. To this end, we 
exemplify the obscured fractions by removing the 
central $5\,$pc, concentrating on the starburst-driven flow. 
Under these conditions, we find that during the starburst, the stirred-up gas as well 
as the randomised clump dynamics
-- mainly in the tens of parsecs distance regime -- can make a significant
contribution to the overall obscured fraction. This can be seen by comparing to the 
observationally derived obscured fractions, which are shown by the horizontal dotted 
and dashed lines in the corresponding colours, which have been adapted from \citet{Buchner_15} 
and \citet{Ricci_15}, respectively.
However, it should be kept in mind that the central cut-out region contains a significant
amount of mass (see Fig.~\ref{fig:mass_budget_time}, blue dotted line and Sect.~\ref{sec:eff_AGN}).
Changing the radius of the inner cut-out region shifts the curves in vertical direction. 

\subsection{Effects on AGN tori}
\label{sec:eff_AGN}

AGN tori are expected to be located within the central few parsecs
region \citep{Jaffe_04,Burtscher_13,Tristram_14,Garcia_Burillo_16,Gallimore_16}.
In our simulations, a very dynamic picture emerges in this
radius regime: spiral patterns build up as a consequence of Toomre instability in the 
first few Myr of the evolution until they break up into clumps. Merging together, they
form a dense disc structure around the central 2-3~pc region, which actively forms stars and 
is often affected by mergers with clumps and tidal arms. After most of the clumps have dissolved
(at an evolution time of roughly 230\,Myr),
a tightly-wound spiral density enhancement remains around the central disc-component, which
slowly decreases in density due to the formation of stars.
Additionally taking this distance regime into account in the 
calculation of the obscured fraction of our simulation (see Sect.~\ref{sec:obs_prop}), we find a 
larger scatter compared to the case excluding the central 5\,pc 
and the values reach obscured fractions of 1 for 
the low column density lines of sight for a significant fraction of the time. 
An enhanced time variation is also found after 
integrating the total gas mass within the 
central 5\,pc sphere (blue dotted line in Fig.~\ref{fig:mass_budget_time}).
As this time variation is caused by clumps 
merging into the central (gas and stellar) density concentration, this indicates that the clumpy 
circumnuclear disc evolution might have a significant influence on the dynamics and  
morphology of AGN tori, as well as BH feeding properties on these variation time scale.
This might be a further indication that tori are time-dependent 
or even transient structures, rather than showing long-term
stability during such active phases of nuclear star formation. 
The disturbances caused by inspiraling clumps might as well cause perturbations of the central disc/torus and form warped discs that are 
often observed in nearby Seyfert nuclei due to their maser emission \citep[e.~g.~][]{Greenhill_96,Greenhill_97,Greenhill_03}.
Central warped discs have been found to significantly change the mid-IR characteristics \citep{Jud_17} as well as partly influence the polarisation 
signal (F.~Marin \& M.~Schartmann, 2017, submitted) of the nuclear gas and dust distribution. 
Infalling clumps (or fragments thereof) merging into the central discs might cause significant gravitational heating
\citep{Klessen_10}, leading to phases of geometrically thick discs or tori 
and might trigger short-duration, nuclear activity cycles. Significant 
impact on the IR emission due to these morphological transformations on parsec scale can 
be expected, resulting in a diversity of AGN tori, as e.~g.~seen in a sample of nearby 
Seyfert galaxies, observed with the MIDI instrument \citep{Burtscher_13}.
Our current spatial resolution is not high enough to test this hypothesis and the results 
should be regarded as tentative, but this should be feasible in near-future 
simulations. 

\subsection{Comparison to published simulations}
\label{sec:comp_sim}

\citet{Wada_09} have investigated models of starbursting circumnuclear discs,
targeted at nearby Seyfert galaxies. Starting with an initially constant 
surface density disc, they find that a gas
distribution in (approximate) equilibrium is obtained within less than 5\,Myr. 
This consists of a puffed-up 
toroidal structure on tens of parsecs scale that surrounds a geometrically thin disc on 
several parsec scale. SN feedback in the latter drives a low density, hot outflow with
a velocity of several 100\,km\,s$^{-1}$.  
They assumed a range of SN rates between $5-500\times 10^{-5}\,\mathrm{yr}^{-1}$
(within a radius of 26\,pc and a height of 2\,pc).
Restricting the analysis of our simulations to the central 30\,pc region, 
we find rates of around $10^{-3}$ SN per year over a time frame of 
roughly 150\,Myr (green line in Fig.~\ref{fig:snr_bin_time}). This is close to 
their upper limit. The main difference between the simulations is the 
distribution of gas, which is concentrated to the central region in our exponential
surface density distribution. Self-consistently following the stellar dynamical 
evolution, we find that the star particles are also very centrally concentrated
(Fig.~\ref{fig:galnuc_partdist}, right column and Fig.~\ref{fig:galnuc_particle_radial}) 
in the evolved state 
of our simulations. Consequently, the SN distribution is much more centrally 
concentrated compared to a spatially random distribution within a thin disc as 
was used in \citet{Wada_09}. 
Reaching out to much larger scale heights in our resulting 
thick stellar disc configuration, our model shows more SN explosions in lower density 
regions off the midplane of the disc. The latter are more effective in  
driving stronger outflows and fountain flow behaviour, compared to the toroidal 
structure found in \citet{Wada_09}. 

Overall, the driving of random motions and outflows due to supernova explosions in our 
simulations can be regarded as a mixture between what has been dubbed as 
{\it peak driving} (explosions happen in high density clumps) and random driving 
(where the SN are randomly distributed within the
computational domain) in idealised simulations \citep{Gatto_15,Walch_15}. 
Stars in our simulations (as well as in nature) form in dense gas clumps. Depending on 
the delay time distribution, a fraction of the SN explosions happen at these locations 
of very high gas densities ({\it peak driving}). These correspond mostly to the high mass progenitor stars 
and to clumps that have suffered very little scattering and merging processes with other 
clumps. Due to the short cooling time, mostly small bubbles are formed that have only
little effect on the evolution of the clump and are barely able to drive significant 
outflows from the central region.
The stellar relaxation through the interaction of clumps allows the SN progenitor stars
to leave the high density regions and explode in a lower density environment. This 
changes the mechanism to {\it distributed} or {\it mixed driving}, which results
in a more efficient coupling of the SN input energy to the low density gas, producing 
and sustaining an outflow perpendicular to the disc plane. This clearly shows the 
necessity of (i) a self-consistent treatment of the dynamical evolution of the stellar
distribution as well as (ii) the implementation of a delay time distribution for the 
SN explosions. 

The role of inflows for the cyclic appearance of nuclear starbursts has been 
investigated by a number of publications.
In a global, very high-resolution simulation of a Milky Way-like galactic disc, \citet{Emsellem_15} study the 
self-consistent formation of a stellar bar that regulates the mass transfer towards the central kiloparsec region. 
Accumulating in ring-like distributions at Lindblad resonances, the gas is able to form stars at the edge of the bar only.
Stellar feedback interaction allows further gas feeding towards the centre within a minispiral. Shear inside the 
bar region inhibits star formation, allowing the transfer of gas through the minispiral and the formation of a 
disc at around 10\,pc distance from the centre, in which clumps form that partly disrupt into long filaments 
due to the tidal interaction with the SMBH or form a nuclear star cluster. Many similarities are found between 
the gas and stellar distribution compared to the Galactic Centre of the Milky Way.
Compared to our circumnuclear disc in isolation, this simulation including the feeding from
larger scales shows a similar behaviour. Whenever gas is able to accumulate, stars are formed, which 
confirms the importance of gravitational instability for the formation of structure in the circumnuclear 
environment. The Milky Way simulation shows
the case of a low mass inflow rate that leads to rather low mass concentrations and episodic star 
formation on short time scale in this very inactive galactic nucleus. 
Our massive gas disc must have either
accumulated over a longer time period or must have been formed by a high mass inflow rate in order to
cause a much stronger starburst and fast, hot outflowing gas.

Concentrating on the central 500\,pc region of the Milky Way (the so-called {\it Central Molecular Zone}, CMZ),
a self-consistent model for star formation cycles based on theoretical arguments and observed properties has been developed 
by \citet{Kruijssen_14}. Within a galaxy-scale gas inflow (e.~g.~bar-driven), acoustic and gravitational instabilities concentrate
the gas until a star formation density threshold is reached. The latter might be environmentally dependent and increased in 
galactic nuclei due to enhanced levels of turbulence, which is consistent with the assumptions of our simulations. 
Their model allows to explain the currently
low star formation rate (SFR) in the CMZ. It seems to be in the slow phase of accumulating gas that limits the rate of star formation. 
This will be followed by rapid gas consumption into stars, once the density threshold for gravitational instability 
is reached.  
Our simulations assume an initial condition that is comparable to the end of the accumulation stage of the model 
found by \citet{Kruijssen_14}  
and gravitational instability is able to occur.
Given the higher accumulated mass, a more intense and a longer starburst follows compared to what is 
found for the Milky Way. This is the phase we concentrate on in this publication and resolve in great detail in space 
as well as time. After roughly 250~Myr, 
our disc returns into the {\it accumulation phase} again with a star formation rate below what is expected 
from the Kennicutt-Schmidt relation (see Fig.~\ref{fig:kenn_schmidt}).
Hence our initial condition and the expected cyclic behaviour is consistent with models including the feeding 
of gas from larger scales.

\subsection{Numerical issues and resolution dependence}

The simulation of circumnuclear discs with sub-parsec resolution is especially demanding, 
as they are located deep within the gravitational potential well of the stellar bulge and 
SMBH and have very high density contrasts.
This not only necessitates very high 
spatial resolution, but the corresponding dynamics and short cooling 
timescales require very short timesteps, making the simulations 
very computationally expensive. As we are interested in simulating 
full activity cycles, long time evolutions of several tens of revolutions of the 
disc are necessary. Another reason for high resolution is to avoid artificial 
fragmentation of the self-gravitating gas. To this end, 
we require that the Jeans length is resolved with at least 4 grid cells, following 
\citep{Truelove_97}. This is achieved first of all by triggering refinement when the Jeans length is 
not resolved well enough (see Sect.~\ref{sec:numhydro}).
To fulfill the Truelove-criterion even at the highest refinement level, the usual artificial pressure floor is used
(Sect.~\ref{sec:numconcepts}).
This only affects the highest densities within the clumpy structures. 
The consequence of using a density threshold for star formation as large as $2\times 10^6\,\mathrm{cm}^{-3}$ is that
for our given maximum resolution, a fraction of the star formation happens within cells that are heated to $10^4-10^5$\,K.
However, these cells are deeply embedded into the high density gas clumps. 

We generally find good convergence of the simulation when increasing the resolution by 
halving the refinement threshold masses by factors of two. The same is true for a simulation
with double the stellar particle mass, e.~g.~converging to the same stellar velocity dispersion,
with differences only arising in the early relaxation phase. 

Using a Cartesian grid, the code conserves linear momentum only. The high resolution
in the fragmenting part of the disc guarantees a reasonable angular momentum conservation (within 10\%) 
in this region of the computational box. In this publication, we are not interested in the evolution
of the smooth outer disc. Including the latter in the angular momentum balance, we find deviations from 
the initial value of up to 50\% at the end of the simulation, which corresponds to roughly 45 orbits 
at a radius of 150\,pc. This is evident in the evolution of the surface density in the outer part of 
the radial range shown in Fig.~\ref{fig:galnuc_radial_plots}b as well as
the changing size of the outer disc (Fig.~\ref{fig:galnuc_dens_evolution}a-d).

\subsection{Limited physics}
\label{sec:lim_phys}

Our simulations start with an (isolated) Toomre unstable, self-gravitating, high density gas disc in approximate
vertical hydrostatic equilibrium.
We show that such an initial configuration allows recovery of many observed properties of nearby circumnuclear discs during
the evolution of the simulation.
Resulting disc masses (gas and stars), the kinematical state, star formation properties and the subsequent driving of a wind are in
reasonable agreement with observed properties.
However, the question remains how such a disc can be build up and stabilised against collapse.
Gas at such high densities cools on a very short time scale. Allowing the cooling down to molecular cloud core temperatures 
would lead to a collapse to a thin (unresolvable) disc within an orbital time scale in our simulations. Numerically, we prevent this 
by setting a minimum temperature of $10^4$\,K, which is thought to replace unresolvable micro-turbulence, corresponding
to a velocity dispersion of the order of $10\,$km\,s$^{-1}$ as well as photoionisation heating once a young stellar population
has been formed.  
As the Toomre stability criterion depends on the gas temperature and kinematical state of the gas, this assumption has 
an important effect on the gravitational stability and allows investigation of well-defined initial conditions. 
In reality nuclear regions are not isolated from larger 
galactic features, but gas is dynamically driven inwards through bars, spirals, etc.
Plunging into the central disc, this accreted gas releases part of its kinetic energy and leads to turbulent 
gas motions within the disc. The strength of this effect depends on the amount of gas present, the mass infall rate and velocity, 
the turbulent driving scale as well as an efficiency factor.
Balancing the energy input with the turbulent dissipation rate, \citet{Klessen_10} find

\begin{eqnarray}
\dot{M}_\mathrm{in} = 0.39\,\mathrm{M}_\odot\,\mathrm{yr}^{-1}\,\left(\frac{0.1}{\epsilon}\right)\,
                      \left(\frac{M_\mathrm{gas}}{10^8\,\mathrm{M}_\odot}\right)\,
                      \left(\frac{\sigma}{10\,\mathrm{km}\,\mathrm{s}^{-1}}\right)^3\,        \\
                      \left(\frac{100\,\mathrm{pc}}{L_\mathrm{d}}\right)\,
                      \left(\frac{160\,\mathrm{km}\,\mathrm{s}^{-1}}{v_\mathrm{in}}\right)^2.
\end{eqnarray}

Using the gas disc mass of our initial condition (M$_\mathrm{gas}=10^8$\,M$_{\odot}$), a gas infall velocity similar to the 
rotation velocity ($v_\mathrm{in}=160$\,km\,s$^{-1}$), a turbulent driving scale similar to the thickness of the 
gas disc ($L_\mathrm{d}=100$\,pc) and a typical efficiency of $\epsilon=10$\% \citep{Klessen_10}, 
the process requires a mass infall rate of the order of 0.39\,M$_{\odot}$\,yr$^{-1}$ 
to reach a velocity dispersion of roughly $\sigma = 10\,\mathrm{km}\,\mathrm{s}^{-1}$. 
Following these simple energy arguments, the resulting 
mass infall rates are in the range of observed infall rates of 0.01-1\,M$_\odot\,\mathrm{yr}^{-1}$ in nearby Seyfert galaxies 
\citep[e.~g.~][]{Storchi_Bergmann_14}. 
This simplistic analysis also shows that the more mass is accumulated by infall and the closer we get 
to our high mass initial state, the more difficult it is to maintain the necessary turbulent motions.
As the reasoning is based on simple energy arguments, high resolution simulations including 
the feeding from larger scales are needed in order to test this hypothesis. This will also lead to 
replenishment of the gas compared to our current closed-box model, allowing either prolongation 
of the star formation episode or initiation of subsequent star formation episodes. 
Whether such a small-scale turbulent pressure floor is equivalent to a thermal pressure is a question
on its own.

Only limited physical mechanisms are applied in our implementation of stellar feedback.
We idealise it in a single event, which we denominate as a supernova explosion.
Each event introduces an energy of $10^{51}$\,erg into the surrounding medium. In reality, massive stars interact with their 
environment in multiple ways. They drive winds of various strengths and morphologies and ionise their surrounding interstellar 
medium. Even the supernova explosions themselves show a variety of energies.
To this end, the feedback modelling we use in this first set of simulations should be regarded as an effective total energy input, summarising
the mentioned effects, that contribute to the scatter around this energy input and add to its uncertainty. Future modelling efforts 
should allow to determine the main effects and to concentrate on the small scale differences.

The clump dynamics and stellar feedback enhance the effective viscosity of the disc and drive gas inwards. 
According to its angular momentum, the gas will finally end up in the accretion disc / torus system and -- after a 
not very well known time delay depending on the effective viscosity -- will lead to the activation of the central nuclear activity.
In this paper we concentrate on the effect of stellar feedback alone. Nuclear activity mainly impacts the central region of the circumnuclear 
disc, possibly driving a high velocity outflow, including some back flow that drives turbulent motions and might lead to the 
formation of high column density toroidal structures \citep[e.~g.~][]{Wada_12,Wada_15}.

\section{Conclusions}
\label{sec:conclusions}

Using high-resolution hydrodynamical simulations with the {\sc Ramses} 
code \citep{Teyssier_02} we investigate the 
interstellar medium in initially marginally stable self-gravitating
circumnuclear discs in galactic nuclei.
We follow the gravitational collapse, the non-linear growth of structure
and the subsequent nuclear starburst phase self-consistently,
including the gravitational potential of a central SMBH and a
galactic bulge (following scaling relations), the self-gravity of the gas disc, 
standard optically thin cooling and a model SN delay time distribution.  
First, concentric rings and spiral-like features form due to Toomre instability 
that -- during the non-linear
evolution of the instability -- break up into clumps
which are able to form stars. Clump-clump merging as well as scatterings lead to the 
formation of a thick gas and stellar disc. During the peak of the nuclear starbursting 
phase, we find that the gas forms a three component structure: (i) An inhomogeneous,
high density, clumpy and turbulent cold disc is 
puffed-up due to an increase in vertical velocity dispersion caused by clump-clump interactions 
and should be observable with ALMA. (ii) It is surrounded
by a geometrically thick distribution of dense clouds and filaments 
above and below the disc midplane which form a fountain-like flow, 
visible in the infrared IFU data. (iii) And a tenuous, hot, ionised
outflow (visible in HST imaging data) leaves the disc perpendicular to its midplane. 
The star formation decreases as a larger and larger fraction of the 
high density gas clumps are used up 
by star formation and have partly expelled their gas by stellar feedback processes.
For the initially marginally stable gas disc around a low mass black hole 
($10^7\,$M$_\odot$) and bulge ($9\times 10^9\,$M$_\odot$) under investigation, the active starbursting phase takes 
of the order of 150-200\,Myr after which it turns into an almost quiescent disc with a low star formation and
outflow rate.
Starting from initial and boundary conditions that are guided by observed
properties of nearby active galactic nuclei, we find good correspondence of many of the simulated stages of evolution with observed 
quantities. A handful of star forming clumps are frequently found in observations of 
nearby circumnuclear discs. The   
nuclear disc masses, star formation rates, supernova rates as well as 
wind mass loss rates derived from integral field unit observations are in reasonable agreement
with the findings of our simulated galactic nucleus. 
In order to achieve this, it is necessary to follow the formation of the stars and 
their subsequent dynamical relaxation process self-consistently and model a delay time 
distribution of the SN feedback, to allow a fraction of the stars to  leave
their parent high density clump. 
Another result of this evolution is a disc of young stars (with a tentative central
nuclear star cluster component) that fits the recently found size-luminosity relation for 
such nuclear stellar discs as well as their observed kinematical state.
The combination of such simulations 
with the large amount of high resolution, 
multi-wavelength, multi-instrument observational data that will be available in 
the coming years will test the scenario described above and
will allow us to build a consistent picture of inflow,
excitation, SF and outflow of the circumnuclear gas. 

\section*{Acknowledgements}

We thank the referee, Fr\'{e}d\'{e}ric Bournaud, for a very positive and encouraging report.
For the computations and analysis, we have made use of many open source software packages, 
including {\sc Ramses} \citep{Teyssier_02}, yt \citep{Turk_11}, NumPy, SciPy, matplotlib, hdf5, h4py. 
We thank everybody involved in the development of these for their contributions.
This work was supported by computational resources provided by the Australian Government through 
NCI Raijin under the National Computational Merit Allocation Scheme, as well as by the 
swinSTAR supercomputing facility at Swinburne University of Technology.
Part of the computations were performed on the HPC system HYDRA of the Max Planck Computing and Data Facility.
MS thanks Bernd Vollmer for helpful discussions. 
MS thanks for support by the SFB-Tansregio TR33 ``The Dark Universe'' and support from
the German Science Foundation (DFG) under grant no.~BU~842/25-1.
JM, MD, and MS acknowledge the continued support of the Australian Research Council 
(ARC) through Discovery project DP140100435.
KW was supported by JSPS KAKENHI Grant Number 16H03959.
LB acknowledges support from a DFG grant within the SPP 1573 ``Physics of the interstellar medium''.




\bibliographystyle{mnras}
\bibliography{literature} 





\bsp	
\label{lastpage}
\end{document}